\documentclass[iop]{emulateapj-rtx4}
\shortauthors{Sekanina \& Kracht}
\shorttitle{Two Long-Period Comets and Resonance-Driven Inward Drifting of
 Aphelion}
\slugcomment{Version \today }

\newcommand{\Msun}{${\cal M}_{\mbox{\scriptsize \boldmath $\odot$}}$}

\begin{document}
%
%
\title{Study of a Potential Relationship Between Two Long-Period Comets and\\Rapid
Inward Drifting of Aphelion Due to Orbital-Cascade Resonance}

\author{Zdenek Sekanina$^1$ \& Rainer Kracht$^2$}
\affil{$^1$Jet Propulsion Laboratory, California Institute of Technology,
  4800 Oak Grove Drive, Pasadena, CA 91109, U.S.A.\\
$^2$Ostlandring 53, D-25335 Elmshorn, Schleswig-Holstein, Germany}
\email{Zdenek.Sekanina@jpl.nasa.gov\\
{\hspace*{2.59cm}}R.Kracht@t-online.de{\vspace{-0.2cm}}}

\begin{abstract}
\noindent
%
We study a potential genetic relationship of comets C/1846 O1~and~C/1973\,D1,~whose~\mbox{apparent}~\mbox{orbital}
similarity~was~tested~by~Kres\'ak~(1982)~only~\mbox{statistically},\,\mbox{using}~the~\mbox{Southworth-Hawkins}\,(1963)\,\mbox{criterion}
$D$.  Our orbit determination for C/1846\,O1 shows its period
was\,$\sim$500\,yr,\,$\sim$30 times shorter than that of C/1973~D1.  Formerly
unrecognized, this incongruity makes the objects'\,common origin~less~likely.
Long-term orbit integration suggests that, if related, the two comets
would~have~to~have~sepa\-rated far away from the Sun (probably $\sim$700~AU)
21~millennia ago and, unlike C/1973~D1, C/1846~O1 would~have~to~have been
subjected to a complex orbital evolution.  Given the chance of encountering
Jupiter to $\sim$0.6~AU some 400~days after perihelion, C/1846~O1
and C/1973~D1~may~have~been~perturbed, during their return in the 15th
millennium BCE, into {\vspace{-0.01cm}}orbits that were, respectively,
smaller and larger than was the parent's, with a net difference of more
than 0.002~(AU)$^{-1}$ in $1/a$.  Whereas C/1973~D1 was on the way to its
1973 perihelion, C/1846~O1 should have been subjected to recurring encounters
with Jupiter, during which the orbital period continued to shorten by
integral multiples of the Jovian orbital period, a process called {\it high-order
orbital-cascade resonance\/}.  While~the~inte\-grated perturbation effect of
C/1846~O1 by Jupiter does not appear to reduce the comet's orbital period~to
below $\sim$1200~yr by the mid-19th~century, we find that orbital-cascade
resonance offers an attractive mechanism for rapid inward drifting of aphelion
especially among dynamically new comets.{\vspace{0.05cm}}
\end{abstract}

\section{Introduction{\vspace{-0.05cm}}}
We recently expounded the relationship between fragmentation and the orbital
properties for two well-known groups of genetically related long-period comets:\
one was the pair of C/1988 F1 (Levy) and C/1988 J1 (Shoemaker-Holt) and the
other was the trio of C/1988 A1 (Liller), C/1996 Q1 (Tabur), and C/2015 F3
(SWAN) (\mbox{Sekanina} \& Kracht 2016, referred to hereafter as Paper~1).
Prior to the discovery of the two groups, a genetic relationship as the
provenance for close orbital similarity among comets was a subject of controversy,
especially in the 1970s and early 1980s.  In a contribution to the debate,
Kres\'ak (1982) corroborated{\vspace{-0.075cm}} and extended Whipple's
(1977) criticism of \"{O}pik's (1971) conclusion on the omnipresence of
groups of related comets.  Kres\'ak argued that there was no compelling
evidence for the existence of a single pair or group of long-period comets
that derive from a common parent, other than the Kreutz system of
sungrazers.{\vspace{0.05cm}}

The approach employed in the pre-1988 debate was always statistical in nature.
In particular, Kres\'ak (1982) used a $D$-criterion, introduced by Southworth
\& Hawkins (1963) in their investigation of meteor streams.  As a function
of differences in the five orbital elements --- the argument of perihelion
$\omega$, the longitude of the ascending node $\Omega$, the inclination $i$, the
perihelion distance $q$, and the eccentricity $e$ --- the $D$-criterion allows
one to express the degree of similarity between two orbits in one-dimensional
phase space.  Objects in orbits of the same spatial orientation that are
identical in size and shape have \mbox{$D = 0$}. The $D$-values of the
genetically related 1988 pair and the 1988--2015 trio, referred to above,
are listed in Table 1.  They never exceed $\sim$0.008, just as the $D$-values
for the Kreutz system's most tightly associated members (such as C/1882 R1 and
C/1965 S1).

The prime subject of Kres\'ak's study was a distribution of $D$-values among
546 comets [a majority extracted from Marsden (1979) and several added] that
arrived at perihelion before the end of 1980 and whose orbital periods exceeded
200 years; the Kreutz sungrazers were excluded.  After finding 38 pairs (and
several chains) of comets with \mbox{$D < 0.3$} and comparing them with
three independent distributions of 546 randomly generated orbits (accounting
in part for observational selection effects), Kres\'ak concluded that the
set of long-period comets exhibited no signs of nonrandom distribution.  In
particular, he judged orbital similarity of the comet pair with the least
$D$-value of 0.084 --- C/1846~O1 (de Vico-Hind) and C/1973~D1 (Kohoutek) ---
not to be statistically significant on the grounds that comparison with the
least $D$-values in the random samples, 0.101--0.120, suggested, on the average,
a $\sim$20\% expectation that this was a random pair as well.{\hspace{0.21cm}}

Unfortunately, Kres\'ak's failure to remove from his statistics the grossly
inferior orbits of comets observed in early times marred his main results, and
he addressed a much more meaningful subset of 438 long-period comets from the
period of 1800--1980 rather inadequately.  Our closer examination of this subset
shows that the minimum $D$-value in the three random samples then moves up to a
range of 0.113--0.134 and there is only an 11\% expectation for C/1846~O1 and
C/1973~D1 being a random pair.  Moreover, if this pair is removed from the set,
the next pair's $D$-value of 0.129 is consistent with an average random sample
with an expectation of 60\%.

\begin{table}[t] 
\vspace{-3.9cm}
\hspace{4.23cm}
\centerline{
\scalebox{1}{
\includegraphics{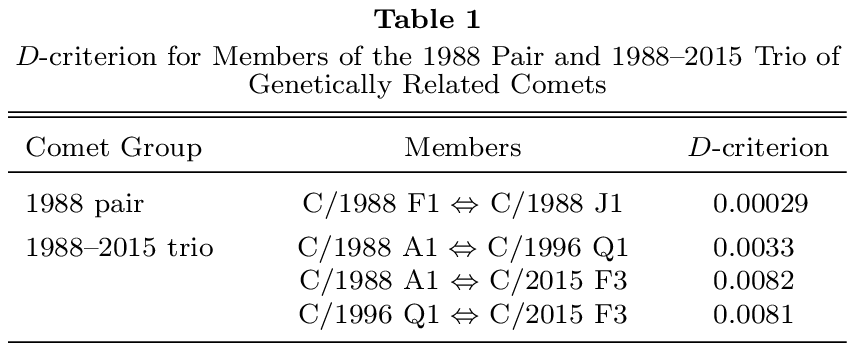}}} 
\vspace{-22.1cm}
\end{table}

Interestingly, in an investigation that extended that~of Kres\'ak (1982),
Lindblad (1985) found that an updated set of long-period comets displayed,
at best, only margin\-ally greater orbital similarity than did random
samples.{\hspace{0.3cm}}

From the statistical standpoint, the pair of C/1846~O1 and C/1973~D1 appears
to be something of an oddball; at first sight, the orbital differences are
not so plainly minute as those of obvious fragments of a common parent (as,
e.g., C/1988~F1 and C/1988~J1; Table~2), yet they are not conforming to a
random distribution so readily as the other fortuitous pairs on Kres\'ak's
(1982) list.  Comparison of the 1846--1973 pair's $D$-value of 0.084
with those in Table~1 suggests that this pair is orbitally bound together
one order of magnitude less tightly than the 1988--2015 trio and two orders
of magnitude less tightly than the 1988 pair.  On the other hand, it should
be emphasized that some members of the Kreutz system, although genetically
related beyond any doubt, have orbits far less similar and their $D$-values
much larger than the 1846--1973 pair.  For example, \mbox{$D = 0.223$} for
the pair of C/1963~R1 (Pereyra) and C/1965~S1 (Ikeya-Seki), both Kreutz
sungrazers with very accurately determined orbits, although classified by
Marsden (1967) as members of different Kreutz subgroups.

\begin{table}[b] 
\vspace{-3.8cm}
\hspace{4.22cm}
\centerline{
\scalebox{1}{
\includegraphics{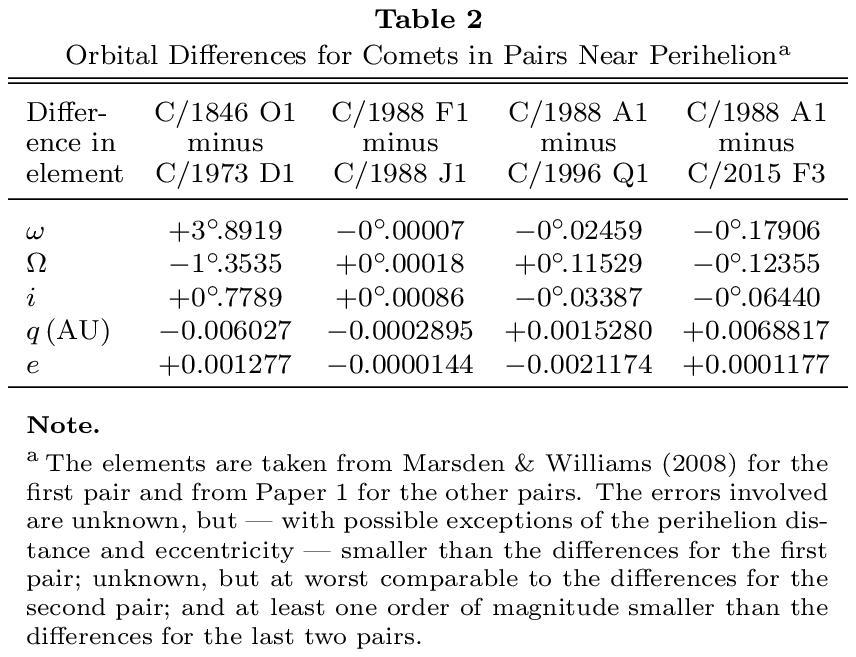}}} 
\vspace{-19.18cm}
\end{table}

In Paper 1 we noted that the 1988 pair (C/1988 F1 and C/1988 J1) was with
high probability a single comet less than one half the orbital period before
discovery, while the parent of the 1988--2015 trio was likely to have split
near the previous perihelion passage.  The process of fragmentation of the
Kreutz system began at least two (Sekanina \& Chodas 2004){\vspace{-0.07cm}}
and possibly many more (Marsden 1989; \"{O}pik 1966) revolutions about the
Sun before the 19th and 20th century clusters were observed.  It appears that
the $D$-criterion, as a measure of orbital similarity, increases with the time
elapsed since the fragmentation event, a trend that is by no means surprising.

By the same token, however, the $D$-criterion proves an unreliable tool in an
effort to investigate a genetic relationship between two particular comets of
unknown histories, and for other than statistical purposes appears to be useless.
Indeed, in meteor astronomy --- for which the $D$-criterion was developed ---
its application is limited to statistical studies only and is therefore fully
justified.

This experience suggests that a much more rigorous approach --- pursued below
--- is required in order to gain insight into the fundamental issue of our
interest:\ {\it Are comets C/1846~O1 and C/1973~D1 genetically related?}

\section{Published Observations and Orbits of\\C/1846 O1 and C/1973 D1}
Comet C/1846 O1 was discovered independently by F.\ de Vico in Rome and by
J.\ R.\ Hind in London on 1846 July 29, some 2 hours apart (de Vico 1846;
Bishop 1852).  The comet was observed astrometrically for about two months,
in September until the 26th only by Argelander (1865) in Bonn.  The orbit in
Marsden \& Williams' (2008) catalog was computed by Vogel (1868); although
the best in existence at this time, the orbit is only a parabola with no
planetary perturbations applied and is clearly inadequate for an in-depth
investigation of the comet's motion.  No physical observations were made,
and the comet's intrinsic brightness published by Vsekhsvyatsky
(1958),\footnote{A so-called absolute magnitude $H_{10}$ is related to
an apparent magnitude $H$ according to a formula \mbox{$H_{10} = H - 5
\log \Delta - 10 \log r$}, where $\Delta$ and $r$ are, respectively, the
comet's geocentric and heliocentric distances; this formula reflects an
assumption that the comet's observed brightness varies as
$\Delta^{-2}r^{-4}$.} \mbox{$H_{10} = 6.2$}, is an estimate based on apparent
magntidues assigned depending on the reported type and size of instruments
large enough or too small to detect the comet.\footnote{For example, describing
the comet on 1846 July 30, Hind remarked that it ``could be just perceived with
the ordinary sea-glass'' (Bishop 1852, p.\,217).  The assigned $H_{10}$ magnitude
of 6.2 is readily obtained by ascribing a limiting magnitude of 9.5 to a glass of
30~mm aperture diameter and a magnification of 8$\times$ at the comet's elevation
of \mbox{25--30}$^\circ$ above the horizon.} 

Comet C/1973 D1 was discovered photographically by Kohoutek (1973a) in
Hamburg-Bergedorf on 1973 February 28 and observed astrometrically for nearly
7~months, for the last time on September 22 by the discoverer (Kohoutek 1973b).
The comet's astrometry is summarized (including the references){\vspace{-0.04cm}}
in the Minor Planet Center's Orbits/Observations Database.\footnote{See {\tt
http://www.minorplanetcenter.net/db\_search}.} Marsden (1973) noticed orbital
similarity with C/1846 O1 from an early parabolic orbit based on a 5.9-days
long arc, but it took much more time to rule out the identity of the two comets
(Marsden 1974).  The currently available high-quality orbit, computed by Marsden,
reveals that the comet has last been near the Sun  some 16\,500~years ago
(Marsden et al.\ 1978).  Brightness estimates{\vspace{-0.04cm}} are the only
physical observations that are available:\footnote{In at least one case the
physical observations that referred to C/1973 E1 (Kohoutek) were mistakenly
assigned to C/1973 D1 (Combi et al.\ 1997).}\,a series obtained visually by
Bortle (1982) before perihelion and a number of approximate total and ``nuclear''
photographic  magnitudes (e.g., Kojima 1973, Kohoutek 1973b) that span almost
the entire period of astrometric observations.  

\begin{figure*}
\vspace{0.15cm}
\hspace{-0.22cm}
\centerline{
\scalebox{0.87}{
\includegraphics{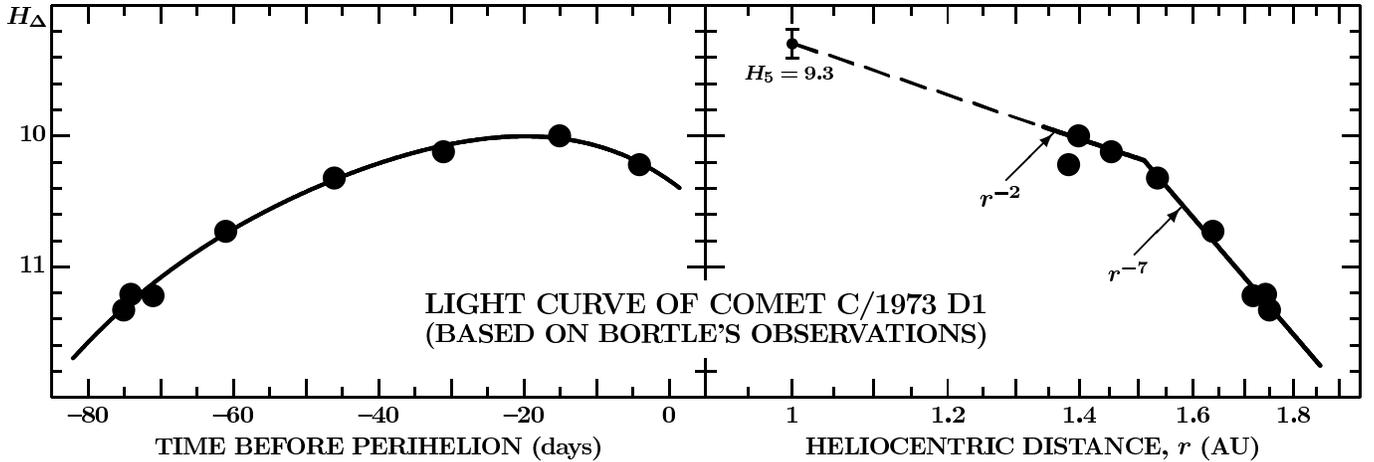}}} 
\caption{Light curve of C/1973 D1 based on the observations by Bortle (1982).
The visual magnitude $H_\Delta$, aperture corrected and normalized to a unit
geocentric distance by an inverse square power law, is plotted against time
(reckoned from perihelion) on the left and against heliocentric distance on
the right.  The fitted polynomial of power 4 (on the left) suggests that the
comet's normalized brightness may have peaked before perihelion and that the
rate of preperihelion brightening with heliocentric distance (on the right)
dropped rapidly around 1.5~AU from the Sun, in late April, when the comet
had $\sim$42 days to go to perihelion.  An extrapolated intrinsic magnitude,
$H_5$, based on the assumption of validity of an inverse square power law
near perihelion amounts to \mbox{9.3$\:\pm\:$0.1}.{\vspace{0.3cm}}}
\end{figure*}

To address the issue of which of the two comets is likely to be intrinsically
fainter, we now examine the light curve of C/1973~D1.  Based on Bortle's (1982)
brightness estimates, spanning more than 70 days, from 1973 March~24 to June 3,
and terminating a few days before perihelion, the light curve is plotted in
Figure~1.  The observed magnitudes $H$ were first corrected for the aperture
of the employed 31.6-cm reflector by subtracting 0.6~mag (Morris 1973) and
thereby standardized{\vspace{-0.03cm}} to a photometric scale of the naked
eye.\footnote{Morris (1973) determined {\vspace{-0.04cm}}an aperture correction
for~\mbox{reflectors} to average $-$0.019~mag cm$^{-1}$, applying a correction
of $-$0.47~mag to reduce Bortle's 31.6-cm reflector magnitudes to a standard
aperture of 6.8~cm (equivalent to $-$0.6 mag for the naked eye).  With reference
to their correspondence, Morris stated that Bortle's own aperture correction
was practically identical.} The corrected magnitudes~$H_{\rm corr}$ were
converted to $H_\Delta$ by normalizing to a geocentric distance of \mbox{$\Delta
= 1$ AU} with an inverse square power law, i.e., \mbox{$H_\Delta = H_{\rm corr} -
5 \log \Delta$}.  No phase correction was applied, because in the entire period
of Bortle's observations the phase angle varied only between 25$^\circ$ and
38$^\circ$.  However,~if reduced to a zero phase angle, the data in Figure~1
could be a few tenths of a magnitude brighter.

The normalized magnitude variations of C/1973 D1 in Figure~1 show no sign of
flare-ups, but a polynomial fit of power 4 suggests that the light curve peaked
about 20 days before perihelion.  However, the last observation, made 4~days
before perihelion, may have been affected by a small elongation from the Sun,
which was only 36$^\circ$.  There was a marked tendency for the light curve
to level off as the comet was approaching perihelion.  Down to a heliocentric
distance $r$ of about 1.5~AU, the comet brightened steeply, approximately as
$r^{-7}$.  If this trend continued, the comet's extrapolated normalized
magnitude at \mbox{$r = \Delta = 1$ AU} would have been about~7.  The rate
at which the brightness was increasing at \mbox{$r < 1.5$ AU} is not at all
well determined, but assuming, conservatively, an inverse square power law,
the extrapolated normalized magnitude at a unit heliocentric and geocentric
distances is \mbox{$H_5 = 9.3 \pm 0.1$} (Figure~1).

An average $r^{-4}$ fit implies an absolute magnitude of \mbox{$H_{10} =
8.2 \pm 0.3$}, 2~mag fainter than C/1846~O1.  However, such comparison is
questionable because C/1846~O1 was discovered nine weeks after perihelion, while
the Bortle light curve refers to the time before perihelion.  The difference
should in fact be still greater, if C/1973~D1 was intrinsically fainter after
perihelion.  If Hind's remark on the marginal appearance of C/1846~O1 in a
sea-glass on 1846 July 30.9 UT, 65 days after perihelion, is indeed interpreted
to imply magnitude 9.5, we find for this comet --- neglecting a small aperture
correction of perhaps $-$0.2~mag --- that \mbox{$H_\Delta = 8.3$} at \mbox{$r =
1.65$ AU} and a phase angle of 35$^\circ$.  On the other hand, C/1973~D1 was
photographed by Kohoutek (1973b) on 1973 August~1.0, 55~days after perihelion,
as an object of total magnitude 15, equivalent to \mbox{$H_\Delta = 13.7$} at
\mbox{$r = 1.59$ AU}.  Since photographic observations underestimate the total
brightness compared to visual data, we need to correct for this effect by
comparing them with Bortle's preperihelion observations.  Although Kohoutek
photographed C/1973~D1 several times between March~24 and June~3, he provided no
magnitudes on those occasions. The only total photographic magnitudes\footnote{See
a website {\tt http://www.minorplanetcenter.net/db\_search}.} from the critical
time span were magnitude~14 by Wood at Woolston on March 22.8, magnitude~15 by
Mrkos at Klet' on April 4.9 UT, and magnitude~14 by Hendrie at Woolston on
April 6.9 UT.  Comparison of the two consistent results with $H_{\rm corr}$ from
Bortle's data yields a photographic-to-visual correction of $-$2.6~mag, which
--- if assumed to be approximately applicable also to Kohoutek's magnitude
scale --- implies for his corrected post-perihelion data point \mbox{$H_\Delta
= 11.1$} at \mbox{$r = 1.59$ AU}, that is, 2.8~mag fainter than C/1846~O1 at
a slightly larger heliocentric distance.  This essentially 3~mag difference,
whose uncertainty is estimated at some $\pm$0.5~mag, suggests that if the two
comets are genetically related, then C/1846~O1 is likely to be the primary and
C/1973~D1 its companion, a conclusion that will serve as a test in our
orbit-evolution modeling.

\section{Strategy of the Present Investigation}
To address our primary objective --- the relationship between the two comets ---
requires a comprehensive examination of the histories of their orbital motion
and an in-depth study of their common parent's most probable motion.  This
general strategy consists of several consecutive steps, dealt with in the
following subsections.

\subsection{Possible Fragmentation Scenarios}
If C/1846 O1 and C/1973 D1 should be fragments of a common parent, their
arrival times at perihelion and the other elements must satisfy certain
conditions depending on the parent's fragmentation time.  One fundamental
property of the groups of genetically related long-period comets (and
nontidally split comets in general) that C/1846~O1 and C/1973~D1 appear
to satisfy is that the primary (intrinsically the brightest and presumably
the most massive) fragment should arrive first.  Based on our experience
that we gained from our study of a pair and a trio of genetically related
long-period comets in Paper~1, the possible scenarios for the timing of
the parent's fragmentation could be divided into three general categories:\
(A)~relatively recently, a fraction of one revolution about the Sun before
their recorded perihelion times in the 19th--20th centurues; (B)~in a
general proximity of perihelion in the previous return; or (C)~substantially
earlier than in the course of the previous return to perihelion.

The first scenario --- a relatively recent event --- is practically ruled out
for two reasons.  One is the unacceptably long period of time --- 127~years ---
between the arrival times of the two comets at perihelion, the other is the
large differences between their other elements, as illustrated in Table~2.
Indeed, adopting the original orbital period of C/1973~D1, computed by Marsden
(Marsden et al.\ 1978), as a first approximation to the parent's orbital period
and ignoring the planetary perturbations, we find (from the equations provided
in Section~2.1.1 of Paper~1) that the fragmentation event would have taken place
at a heliocentric distance of about 127~AU, some 110~years after the previous
perihelion passage or $\sim$16\,400~years ago.  Thus, the assumption of a
fragmentation scenario of category~(A) results in a scenario of category (B).

On the same premise of the parent moving in an orbit with a period of
$\sim$16\,500 years, scenario of category (B) is self-consistent.  In particular,
as shown below, it satisfies the basic relationship between the separation
velocity and the difference of 127~years in the two comets' arrival times,
dictated by Equation~(2) of Paper~1.  Replacing the component of the separation
velocity along the orbital-velocity vector, $\Delta\!V$, with the statistically
averaged separation velocity, $\langle V_{\rm sep} \rangle$ [see the text in
Paper~1 near Equation~(26)], we find on these assumptions for
the presumed pair of C/1846~O1 and C/1973~D1:
\begin{equation}
\langle V_{\rm sep} \rangle = 0.36 \, (r_{\rm frg}/q)^{\frac{1}{2}} \! ,
\end{equation}
where $r_{\rm frg}$ is the heliocentric distance at the fragmentation time
(in AU) and \mbox{$q = 1.382$ AU} is the perihelion distance; the separation
velocity comes out to be in m~s$^{-1}$.  This equation suggests a plausible
submeter-per-second separation velocity near perihelion and a still acceptable
separation velocity of just below 3~m~s$^{-1}$ at the 127~AU from the Sun,
mentioned above.

It appears that the 127-year gap between C/1846~O1 and C/1973~D1 is consistent
with the same straightforward explanation that accounted for the 1988-2015 trio
in Paper~1, and that therefore the 1846-1973 pair is, too, genetically related.
There are only two potential problems:\ (i)~a major difference of more than
3$^\circ$ in the argument of perihelion between the two apparent fragments of
the same parent and (ii)~the premise that the parent --- and therefore also
C/1846~O1 --- had an original orbital period of 16\,000 to 17\,000~years,
close to that of C/1973~D1; both points are so critical to the relationship
issue that an in-depth investigation of the orbital motion of C/1846~O1 was
absolutely indispensable.

\subsection{Improving the Orbit for C/1846 O1}
Because the orbit derived by Vogel (1868) is so highly unsatisfactory, our goal
was to compute an improved solution from scratch, using Vogel's references to
the publications with the original astrometric positions.  We considered it
desirable to replace these with new positions based on comparison-star
positions from the {\it Hipparcos\/} or {\it Tycho-2\/} catalogs.\footnote{The
search facilities are available at the following websites:
{\tt http://\,www.rssd.esa.int/index.php?project=HIPPARCOS\&page= hipsearch} for
the {\it Hipparcos\/} (and the original {\it Tycho\/}) catalog and
{\tt http://vizier.u-strasbg.fr/viz-bin/VizieR-2?-source=I/259/ tyc2\&-out.add=}
for the {\it Tycho-2\/} catalog.}  Our effort was first directed toward
ascertaining, for each published observation, the comet's offsets in right
ascension and declination from the comparison star as well as toward identifying
the star.

Vogel (1868) collected 40 astrometric observations that were made at nine
sites.  Consulting the original references, we soon found that of the 11
positions measured by Hind at Bishop's Observatory in the Regent's Park section
of London (Bishop 1852), eight were not obtained by micrometric comparison with
a field star, but by reading the circles of the equatorial (referred to in the
publication as ``instrumental positions'').  We excluded these from the data
set because of their inherent low accuracy.  In addition, because of incomplete
information available, we could neither determine the offsets from the comparison
star measured by J.\ Challis (Hind 1846) at Cambridge (U.K.) on July~30 nor
identify the comparison star used by E.\ J.\ Cooper (Graham 1846) at Markree
on August~31.  For the remaining 30~observations we were able to recover both
the comparison stars and the offsets.  In addition, we also were able to
identify, in the {\it Tycho-2\/} catalog, comparison stars for two observations,
for which Vogel was lacking information and did not include them in his data
set.  We thus ended up with a total of 32 re-reduced and updated astrometric
positions for our orbit determination.
%

An {\it EXORB7\/} code, written by A.\ Vitagliano and in possession of the second
author, was employed to carry out the computations.  The code accounts for the
perturbations by the eight planets, by Pluto, and by the three most massive
asteroids; and it applies a differential least-squares optimization procedure
to compute the orbital elements.  We began by fitting all 32 observations, of
which three, found to be fundamentally incorrect, leaving residuals in excess
of 80$^{\prime\prime}$, were immediately discarded; comparison with Vogel's
(1868) paper showed that these were the same observations that gave similarly
unacceptable residuals from his preliminary orbit.  An elliptical solution was
then found that satisfied the remaining 29 positions with a mean residual of
$\pm$10$^{\prime\prime\!}$.9, but showed that one observation left a residual
exceeding 3$\sigma$.  Next we applied progressively tighter rejection~cut\-offs
from 30$^{\prime\prime}$ down to 6$^{\prime\prime}$, paying particular attention
to variations in the osculating orbital period as a function of the rejection
cutoff.  The results, exhibited in Table~3, suggest that the osculating orbital
period $P_{\rm osc}$ was nominally always between 350~yr and 700~yr,  that
\mbox{250 yr $< P_{\rm osc} <$ 1000 yr} at 1$\sigma$, and that \mbox{$P_{\rm
osc} < 1600$ yr} at 3$\sigma$.  This result is significant for two reasons:\
one, it shows that a parabola, as employed by Vogel (1868), is an unacceptable
approximation; and, two, indicates that this comet revolves about the Sun
$\sim$10 or more times in the same period of time in which C/1973~D1 makes
just a single revolution.  This major discrepancy and its implications will
be addressed in greater detail in Section~3.4.

\begin{table}[t] 
\vspace{-4.13cm}
\hspace{4.2cm}
\centerline{
\scalebox{1}{
\includegraphics{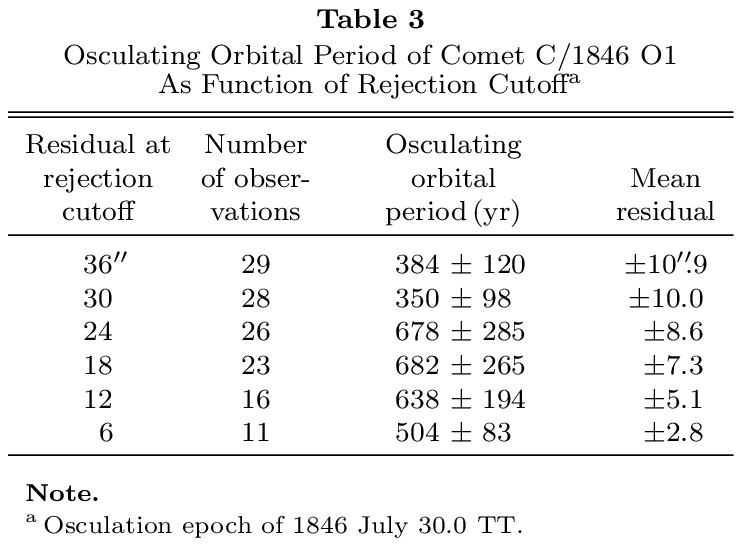}}} 
\vspace{-20cm}
\end{table}

Table 4 offers our preferred orbital solution for comet C/1846 O1, computed for a
standard 40-day epoch of osculation at a rejection cutoff of 6$^{\prime\prime}$.
Relative to Vogel's (1868) orbit, we note --- besides the major deviation from a
parabola --- significant differences in the other elements as well, amounting to
about 10 times the mean error:\ the perihelion time 1.4~days earlier, the argument
of perihelion 1$^\circ\!$.3 lower, the longitude of the ascending node and the
inclination (reckoned relative to the normal to the ecliptic plane) both more
than 0$^\circ\!$.2 higher, and the perihelion distance more than 0.02 AU smaller.
The distribution of residuals from all 32 observations is presented in Table 5;
the rejected observations are parenthesized.

\begin{table*}[t] 
\vspace{-3.8cm}
\hspace{-0.55cm}
\centerline{
\scalebox{1}{
\includegraphics{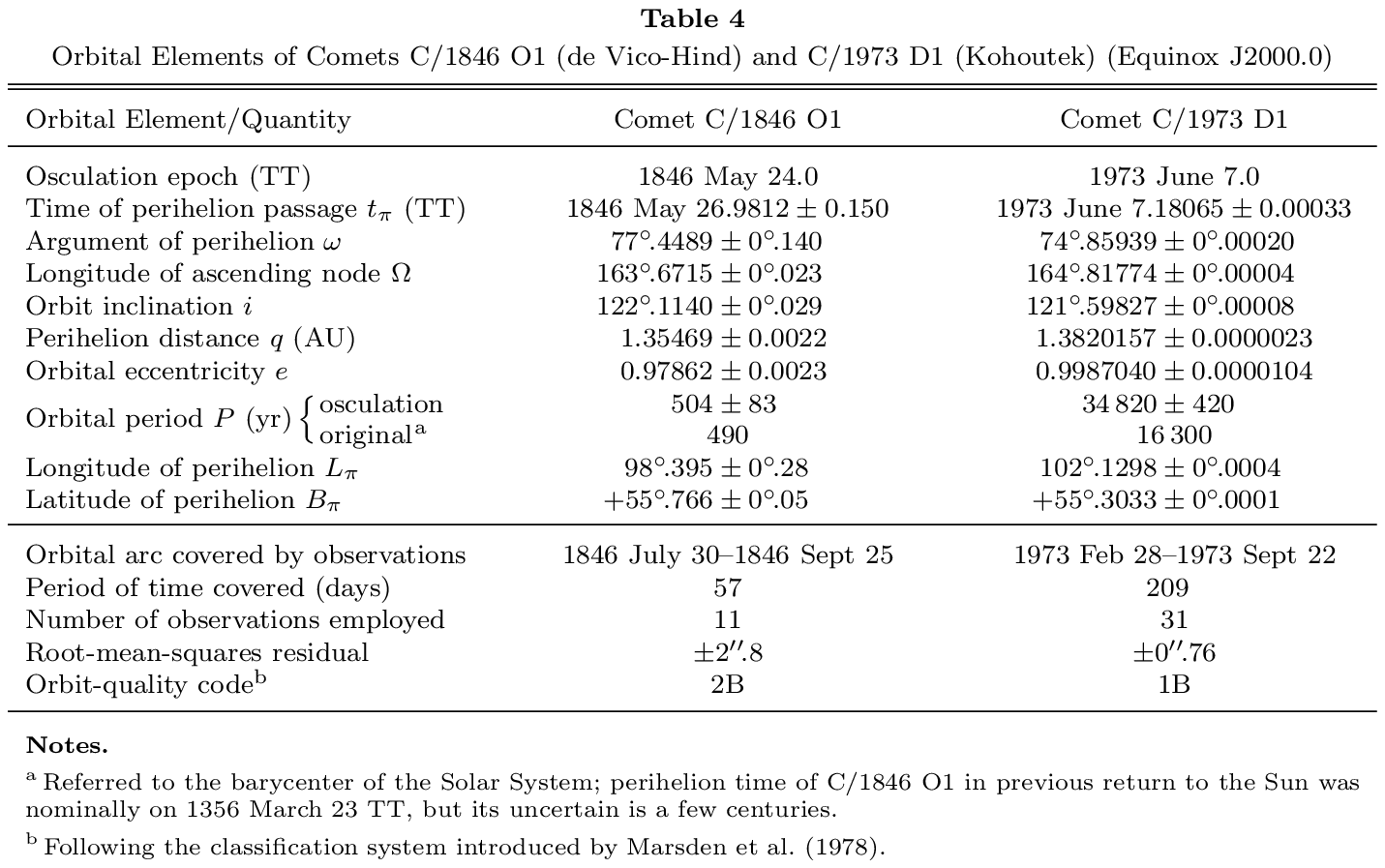}}} 
\vspace{-16.3cm} 
\end{table*}

\subsection{New Orbit for C/1973 D1}
Although we considered taking Marsden's definitive orbit based on 38
observations (Marsden et al.\ 1978; also Marsden \& Williams 2008) as a
starting set of elements for further computations, we eventually decided
to recompute the orbit for two reasons.  One, we preferred a very tight
rejection cutoff for the residuals and were uncertain of what cutoff was
applied by Marsden in his solution.  More importantly, we were bent on
estimating and/or constraining the magnitude of the nongravitational
effects on the motion of the comet, given its feeble intrinsic brightness
(Figure 1), for which we had to get involved with fitting the observations
anyway.  We collected 42~astrometric observations from the Minor Planet
Center's database (Section~2) and from the first orbital run we established
that the residuals from four observations exceeded 3$^{\prime\prime}$, which
thus was the rejection cutoff chosen by Marsden.  We tested solutions at the
rejection cutoffs of 2$^{\prime\prime\!}$.5 (with 34~observations surviving)
and at 2$^{\prime\prime}$ (with 31~observations), and adopted the latter
solution for further investigation of the comet's orbital evolution.
This set of elements is presented in Table~4, while the residuals from all
42~observations are in Table~6, the rejected ones again parenthesized.  A high
degree of similarity between this orbit and that by Marsden is apparent from
the small differences in the individual elements, but thanks to the tighter
rejection cutoff, the mean errors of our orbital elements are nearly a
factor of two lower than Marsden's (Marsden et al.\ 1978).  The differences
in the sense ``Marsden orbit minus orbit in Table~4'' and the mean errors of
the Marsden orbit are:\ +0.00079$\:\pm\:$0.00065~day in the perihelion time,
+0$^\circ\!$.00045$\:\pm\:$0$^\circ\!$.00040 in the argument of perihelion,
+0$^\circ\!$.00001$\:\pm\:$0$^\circ\!$.00007 in the longitude of the ascending
node, $-0^\circ\!$.00010$\:\pm\:$0$^\circ\!$.00017 in the inclination,
+0.0000030$\:\pm\:$0.0000045 AU in the perihelion distance, and
+0.0000192$\:\pm\:$0.0000182 in the eccentricity.  These differences appear to
be comparable to the mean errors of the Marsden set of elements.  On the other
hand, for five of the six elements the differences are clearly larger than the
mean errors of our set of elements in Table~4; on the average the ratio of the
difference to the mean error is 1.55.

We also investigated the distribution of residuals from the solutions with the
rejection cutoffs of 3$^{\prime\prime}$ and 2$^{\prime\prime}$ and found that
the residuals differed systematically by up to 0$^{\prime\prime\!}$.6 in right
ascension and by up to 0$^{\prime\prime\!}$.3 in declination.  Since these
residuals reflect differences of 1.55 times the mean error of the elements, we
considered as acceptable only systematic residuals between the two sets of up
to 0$^{\prime\prime\!}$.2 in right ascension and up to 0$^{\prime\prime\!}$.1
in declination, equivalent, on the average, to deviations in the orbital
elements of up to 0.5 their mean errors.

We were now ready to investigate the effects of the nongravitational acceleration,
employing the standard Style II formalism introduced by Marsden et al.\ (1973).
We examined separately the effects due to a radial component (with an amplitude
of $A_1$ at 1~AU from the Sun), a transverse component (with an amplitude $A_2$),
and a normal component (an amplitude $A_3$).  We obtained specific orbital
solutions by forcing two different magnitudes of the nongravitational
accelerelation, found that the peak residuals varied in proportion to the
acceleration's amplitude, and concluded that each of the three components made
the fit unacceptable by increasing the systematic residuals beyond the allowed
levels unless
\begin{eqnarray}
|A_1| & < & 0.45 \times 10^{-8}\,{\rm AU\:day}^{-2}, \nonumber \\
|A_2| & < & 0.15 \times 10^{-8}\,{\rm AU\:day}^{-2}, \nonumber \\
|A_3| & < & 0.10 \times 10^{-8}\,{\rm AU\:day}^{-2}.
\end{eqnarray}
The components of the actual nongravitational acceleration affecting the orbital
motion of C/1973~D1 must satisfy these conditions in order not to worsen the
distribution of residuals beyond the acceptable levels in right ascension and
declination as defined above.\footnote{In addition, the limit on the
normal component is supported by an orbital solution that incorporated $A_3$
as a variable and resulted in \mbox{$A_3 = (+0.22 \pm 0.18) \times 10^{-8}\,$AU
day$^{-2}$}.}

\subsection{Orbital Properties of C/1846 O1 and C/1973 D1, and Their
Implications}
Perfunctory inspection of Table 4 reveals immediately that C/1846~O1 and
C/1973~D1 arrived at their perihelia at nearly the same time of the year;
while this fact implies that their paths over the sky must have been rather
similar (given that the other elements are also very much alike), this
coincidence is from the standpoint of the two comets' potential relationship
inconsequential.

Our orbital solution for C/1973 D1 in Table 4 confirms that the comet's
previous return to perihelion occurred some 16\,000 to 17\,000 years ago.
Even though the mean error in our solution is almost a factor of two lower than
in Marsden et al.'s (1978) --- $\pm$420~yr vs $\pm$750~yr --- the difference
in the nominal original orbital period is only 220~yr.  On the other hand,
the discrepancy between our and Vogel's (1868) sets of orbital elements for
C/1846~O1 is quite significant, as already pointed out in Section 3.2.  It
should be noted that while the worrisome difference between C/1846~O1 and
C/1973~D1 of 3$^\circ\!$.9 in the argument of perihelion now dropped to merely
2$^\circ\!$.6, the difference in the perihelion distance increased from
0.006~AU to 0.027~AU.  The most striking finding in Table~4 is the orbital
period for C/1846~O1.  Although the quality of the set of elements for this
comet is markedly inferior to that of C/1973~D1, it is extremely unlikely
that the orbital period of C/1846~O1 exceeded $\sim$1600~yr regardless of
the adopted rejection cutoff for the residuals, as is documented by Table~3.
This upper limit on the orbital period is still one order of magnitude
shorter than the original orbital period of C/1973~D1.  The poorly
determined nominal orbital period of C/1846~O1 in Table~4 is more than
30 times shorter than that of C/1973~D1.

\begin{table}[t] 
\vspace{-3.7cm}
\hspace{4.22cm}
\centerline{
\scalebox{1}{
\includegraphics{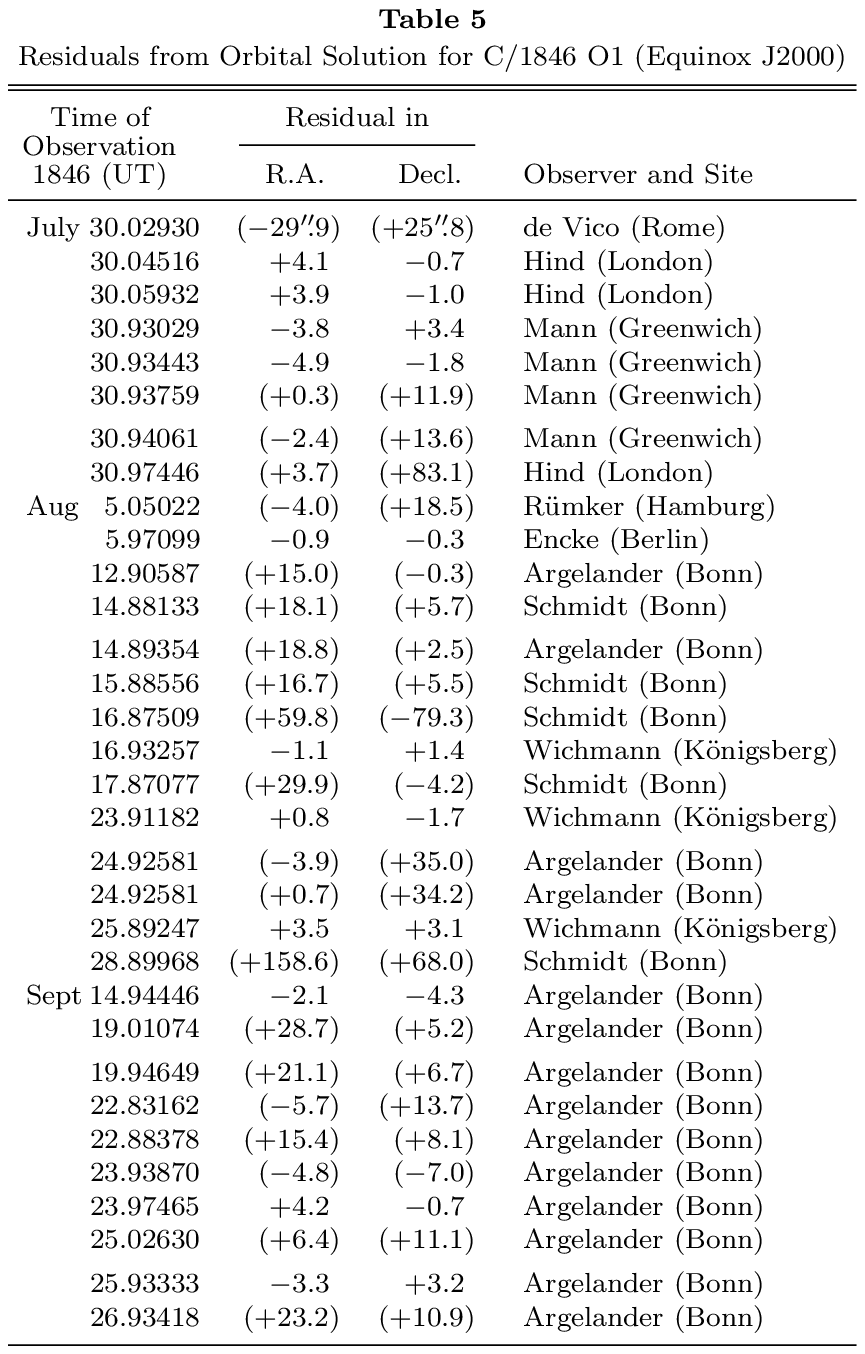}}} 
\vspace{-12.7cm}
\end{table}

\begin{table}[t] 
\vspace{-3.83cm}
\hspace{4.21cm}
\centerline{
\scalebox{1}{
\includegraphics{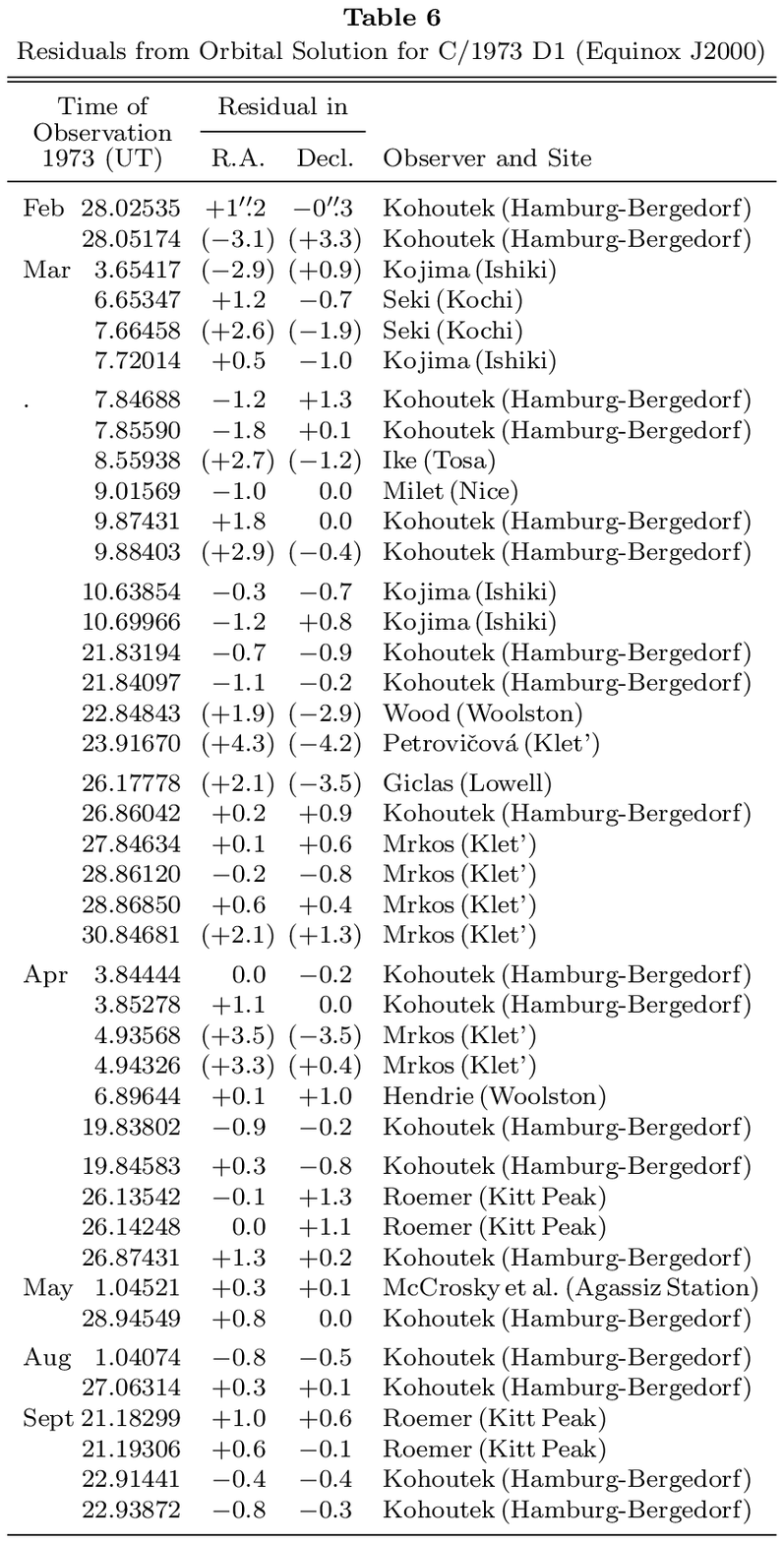}}} 
\vspace{-8.5cm} 
\end{table}

Under these circumstances, a temporal gap of 127~yr between the two comets
is inconsequential and the promisingly looking straightforward explanation
of their genetic relationship, expounded in Section  3.1, is invalidated,
as it is trivial to show that the effects of splitting cannot alone insert
fragments into orbits as different as to have the orbital periods of,
respectively, 500~yr and 16\,300~yr.  Indeed, a differential velocity along
the orbital-velocity vector at perihelion 1.38~AU from the Sun (the only
part of the parent's orbit presumably shared by such fragments before
splitting) comes out in this case to be about 180~m~s$^{-1}$ and it would
not drop below 70~m~s$^{-1}$ even in an extreme case of C/1846~O1's orbital
period of 1600~yr.  Since, in addition, a statistically averaged~\mbox{\it total\/}
separation velocity is $\pi$ times higher than its orbital-velocity component
(Section 2.2.1 of Paper~1), the typical separation velocities implied, some
\mbox{220--560}~m~s$^{-1}$, are fully two to three orders of magnitude higher
than the known separation velocities of the split comets (which are submeter-
to meter-per-second; e.g., Seka\-nina 1982).  Accordingly, C/1846~O1 and
C/1973~D1 must already have been separate objects in the previous return
to perihelion 16\,000--17\,000~years ago, and if they are fragments of a
common parent comet at all, it must have broken up substantially earlier
than in the course of the previous return to perihelion --- invoking the
fragmentation category C in Section 3.1.

\begin{figure}[t] 
\vspace{0.15cm}
\hspace{-0.2cm}
\centerline{
\scalebox{1.05}{
\includegraphics{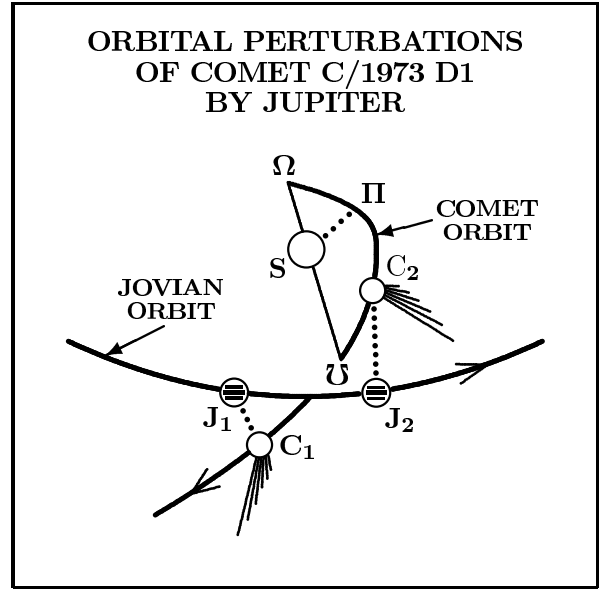}}} 
\vspace{0.02cm}
\caption{Schematic representation of the orbital orientations of comet C/1973~D1
and Jupiter to assess effects of the planet's gravity pull on the comet.  We
depict the Sun (S), the comet's perihelion point ($\Pi$), the nodal line with
the positions of the ascending node ($\Omega$) and the descending node ($\mho$),
and two pairs of relative positions of the comet and Jupiter.  When the comet
is at C$_1$ and Jupiter at J$_1$, the planet's gravity pulls the comet in the
general direction against its motion, thus decreasing its orbital velocity
and period, whereas when the comet is at C$_2$ and Jupiter at J$_2$, the
planet's gravity pulls the comet in the general direction of its motion,
increasing its orbital velocity and period.  Since the Jovicentric distance
in the relative positions of type 1 is typically smaller than in the relative
positions of type 2, the braking effect of the planet's pull is usually
stronger.{\vspace{0.4cm}}}
\end{figure}

Another noteworthy property of the two orbits in Table~4, especially of the
orbit of C/1973~D1, is the position of the line of nodes.  The comet's passage
through its ascending node 122 days before perihelion at a heliocentric
distance of 2.19~AU is of no particular interest, but the passage through its
descending node 274~days after perihelion at a heliocentric distance of 3.74~AU
is significant because the comet is then not too far from the Jovian orbit,
whose heliocentric distance at that longitude is 4.98~AU.  Close encounters
of the comet with the planet, down to $\sim$0.6~AU, are in fact possible
especially around 100~days past the passage through the descending node, when
the comet is south of the planet's orbit plane and moving away from it.  If the
planet is in that part of its orbit at the time, its gravity is pulling the
comet in the direction opposite the motion into an orbit of shorter period, as
schematically depicted in Figure~2 (a configuration \mbox{J$_1$--C$_1$}).  The
Jovian gravity can also pull the comet into an orbit of longer period (a
configuration \mbox{J$_2$--C$_2$}), but the magnitude of such an effect is
generally smaller because of a larger Jovicentric distance of the comet.

The uncertainties in the orbital motions of C/1846~O1 and C/1973~D1, especially
in their orbital periods, prevent us from investigating any particular scenario
of a potential genetic relationship between the two comets.  Instead, we examine
whether {\it such a relationship could in principle be possible\/}.

The orbit of C/1973~D1 in Table~4 is suitable as a starting point to undertake
this task because it is of adequate quality for this purpose and supplies
credible evidence that the potential fragmentation did not occur more recently
than 16\,000--17\,000~years ago.  This conclusion is independent of the fact
that, if related, C/1973~D1 would be a companion to C/1846~O1.  In addition,
the tight limits on the magnitude of the nongravitational acceleration in the
orbital motion of C/1973~D1 [relations (2)] rule out its major effect on the
results; accordingly, we will be working with the gravitational orbit in Table~4.

\begin{figure} 
\vspace{0.2cm}
\hspace{-0.19cm}
\centerline{
\scalebox{0.68}{
\includegraphics{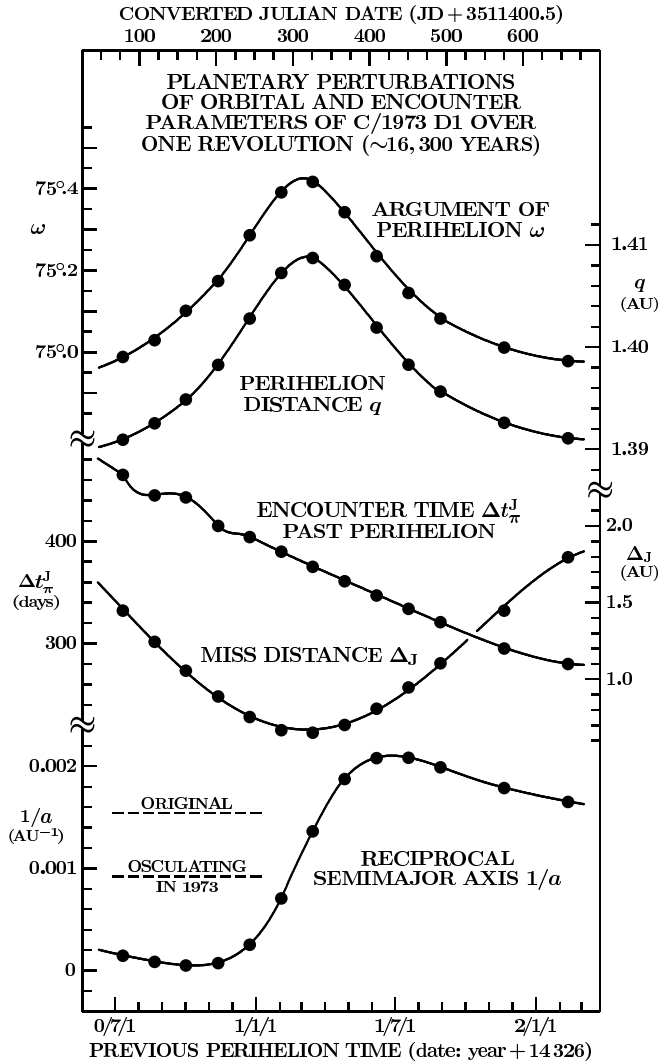}}} 
\caption{Plot, against the varied perihelion time of C/1973~D1 in its previous
return 16\,300~yr ago, of the comet's perturbed orbital elements and
Jovian-encounter parameters, from the top down:\ the argument of perihelion
$\omega$; the perihelion distance $q$; the encounter time,{\vspace{-0.02cm}}
measured as a temporal distance of closest approach to Jupiter from the comet's
perihelion time, $\Delta t_{\pi}^{\rm J}$; the miss distance at the time of
closest approach, $\Delta_{\rm J}$; and the reciprocal semimajor axis, $1/a$.
The osculating $1/a$ at the 1973 perihelion and the original $(1/a)_{\rm orig}$
are also shown.  The uncertainty in the perihelion time is $\pm$420~yr, but,
given its choice, the errors in the plotted quantities are smaller than
{\vspace{-0.045cm}}the size of the symbols.  The quantities $\omega$,
$\Delta t_\pi^{\rm J}$, and $1/a$ have their scales on the left, $q$ and
$\Delta_{\rm J}$ on the right.  Note that year 1 is $-$14\,325 and that the
Julian dates are negative by more than 3.5 million days.{\vspace{0.4cm}}}
\end{figure}

\subsection{Integration of Motion of C/1973 D1 to Previous Perihelion and
Encounter with Jupiter}
The next step in our investigation was the integration of the motion of C/1973 D1
back in time to the previous passage through perihelion, employing an extended
version of the standard JPL DE421 ephemeris.  Since we dealt with time
intervals of up to $\sim$20\,000~years, we first conducted thorough tests of
truncation errors by comparing results of integration to 15 decimal places with
those to 17 decimal places --- the actual precision employed in our computations.
In the absence of close encounters with Jupiter the truncation errors remained
in a subsecond range over the entire period of time.  In the presence of four
encounters, the differences between the two precision limits amounted to about
2~seconds after integration over 10\,000~yr, 30~minutes after 15\,000~yr,
and 1.5~days after 20\,000~yr, so the precision to 17~decimal places assured
us that errors did not exceed 0.02~second after 10\,000~yr, 20~seconds after
15\,000~yr, and 1300~seconds after 20\,000~yr.  We were confident that
truncation should not affect our results appreciably, except when at least
four encounters and integration over $\sim$20~millennia were involved; we
return to this issue in Section 6.
%
%

The reciprocal original semimajor axis of C/1973~D1 derived from the orbital
elements in Table~4 comes out to be
\mbox{+0.0015555$\:\pm\:$0.0000072}~(AU)$^{-1}$.  Direct integration of the
nominal orbit from 1973 back in time resulted in a perihelion time of April~18
of the year $-$14\,321 or 14\,322~BCE.  Since perihelion was also the osculation
epoch of the orbital set, the solution includes the post-perihelion planetary
perturbations of the comet's motion in that return.  The comet did not get
closer to Jupiter than 4.556~AU (10~days after perihelion).  Because the
mean error in the orbital period is $\pm$420~yr, the probability that the
nominal case represents the real situation is very close to zero, which justifies
the search for a range of plausible solutions, primarily those that involve
an encounter with Jupter (Section 3.4).

We addressed this issue by varying slightly the 1973 orbit eccentricity of
C/1973~D1 and by examining~the changing circumstances during the comet's previous
return to perihelion, including its motion relative to Jupiter.  We discovered
that the most significant orbital changes occurred in a range of adopted
eccentricities from \mbox{$e_{\rm nominal} + 0.00000028$} to \mbox{$e_{\rm
nominal} + 0.00000042$}, which differ from the nominal value by 2.7\% to 4\%
of its mean error and are perfectly tolerable deviations.  The major results,
presented in Figure~3, include three orbital elements with the planetary
perturbations integrated down to perihelion --- the argument of perihelion
$\omega$, the perihelion distance $q$, and the reciprocal semimajor axis $1/a$
(all at the osculation epoch of perihelion) --- and two Jovian-encounter
parameters --- the time of closest approach reckoned from the comet's perihelion
time, $\Delta t_\pi^{\rm J}$, and the minimum distance from the planet,
$\Delta_{\rm J}$.  They are plotted against the comet's perihelion time, which
ranges from $-$14\,326~June~1 to $-$14\,324~April~1, thus covering a period of
22~months.

Inspecting three of the curves in Figure 3 --- those for $1/a$, $\Delta t_\pi^{\rm
J}$, and $\Delta_{\rm J}$, we note the prominent variations in $1/a$ that show two
distinct extremes referring to the perihelion times 264~days, or nearly 9~months,
{\vspace{-0.03cm}}apart.  A $1/a$ maximum of +0.002095~(AU)$^{-1}$ that occurs
341~days after perihelion at a minimum Jovicentric distance of 0.874~AU refers
to a perihelion time of $-$14\,325~June~27, whereas a $1/a$ minimum of
+0.000049~(AU)$^{-1}$ that takes place 440~days after perihelion at a
minimum~Jovicentric distance of 1.036~AU refers to a perihelion time of
$-$14\,326~October~6.\footnote{Closest approach possble, at a minimum
Jovicentric distance of 0.649~AU at 379~days after perihelion and referring to a
perihelion time of $-$14\,325~March~6, essentially coincides with an inflection
point of the $1/a$ curve in Figure 3.}  The comet's motion was integrated back
in time; when reckoning the $1/a$ variations in the forward direction, the $1/a$
maximum refers to the case of a peak acceleration of the comet's orbital motion
(a maximum increase in the orbital period), while the $1/a$ minimum to that of
a peak deceleration (a maximum decrease in the orbital period).  We may perceive
the two extremes as belonging to two hypothetical objects, H$_1$ and H$_2$, and
formulate our findings as follows:\\[-0.23cm]

(1a) A hypothetical object H$_1$, moving in the post-encounter orbit of comet
C/1973~D1 and having closest approach to Jupiter 341 days after perihelion
(which took place on $-$14\,325~June~27),{\vspace{-0.04cm}} had at
perihelion~a~pre-encounter \mbox{$1/a = +0.002095$ (AU)$^{-1}$}.\\[-0.23cm]

(1b) A hypothetical object H$_2$, moving in the~post-encounter orbit of comet
C/1973~D1 with~closest~approach to Jupiter 440 days after perihelion (occurring
on $-$14\,326~October~6),{\vspace{-0.03cm}} had at perihelion a pre-encounter
\mbox{$1/a = +0.000049$ (AU)$^{-1}$}.  Thus, H$_2$ passed through its perihelion
264~days earlier than H$_1$ and reached the point of closest approach to Jupiter
\mbox{$264\!+\!341\!-\!440\!=\!165$}~days earlier than H$_1$.  Before the
encounters, H$_2$ was moving ahead of H$_1$ in an orbit that was very similar
to, but slightly more elongated than, the orbit of H$_1$.  The difference
between the two objects in the pre-encounter value of $1/a$ equals
\mbox{$+0.002095 \!-\!  0.000049 \!=\! +0.002046$}~(AU)$^{-1}$ near perihelion,
in the sense H$_1 -$\,H$_2$.\\[-0.23cm]

Next we propose a scenario:\\[-0.23cm]

(2a) Let object H$_1$ be identical with comet C/1973~D1; its pre-encounter
\mbox{$1/a \!=\! +0.002095$ (AU)$^{-1}$} at perihelion, whereas its post-encounter
$1/a$ far from the Sun~\mbox{became} eventually equal to \mbox{$(1/a)_{\rm fut} =
+0.0015555$~(AU)$^{-1}$}.\\[-0.22cm]

(2b) Let a third object, H$_3$, have its Jovian encounter at the same time as
H$_2$, but before the encounter it was moving in the orbit of H$_1$ rather than
H$_2$.\\[-0.22cm]

(2c) Let object H$_3$ be identical with comet C/1846~O1, so that both comets
were moving along the same pre-encounter orbit.\\[-0.22cm]

(2d) Let the gap of 264 days between C/1846~O1 and C/1973~D1 be a product of
their parent's fragmentation some time before they reached perihelion in the
course of the 15th millennium BCE.\\[-0.22cm]

This scenario allows us to arrive at the following three conclusions:\\[-0.23cm]

(3a) Near perihelion, the pre-encounter $1/a$ of{\vspace{-0.03cm}} both C/1846~O1
and C/1973~D1 {\vspace{-0.03cm}}equaled +0.002095~(AU)$^{-1}$ and their orbital
period $\sim$10\,400~yr.\footnote{Here we neglect a slight effect on $1/a$
due to the separation velocity of C/1846~O1 and C/1973~D1,{\vspace{-0.04cm}}
a product of their parent's fragmentation; its magnitude is estimated at
$\sim$0.000001~(AU)$^{-1}$.{\vspace{0.04cm}}} Far from the Sun on the way to
perihelion the comets' orbital period was equal to $\sim$7160~yr. The leading
position of C/1846~O1 suggests that it was the primary fragment, C/1973~D1 was
the companion (cf.\ the comet groups in Paper~1).\\[-0.24cm]

(3b) \mbox{The post-encounter $1/a$ of comet C/1846~O1 was}
\mbox{greater than $1/a$ of comet C/1973~D1 by an amount ap-}{\vspace{-0.03cm}}
proximately equal to 0.002046~(AU)$^{-1}$, the difference be-{\vspace{-0.03cm}}
\mbox{tween the pre-encounter\,$1/a$\,values of +0.000049\,(AU)$^{-1}$}{\vspace{-0.03cm}}
\mbox{and +0.002095\,(AU)$^{-1}$. Since the post-encounter}~$1/a$~of{\vspace{-0.03cm}}
C/1973~D1 eventually equaled +0.0015555~(AU)$^{-1}$~(with the orbital period of
16\,300~yr), it
follows that the post-encounter $1/a$ of C/1846~O1 should have amounted
{\vspace{-0.03cm}}to about
\mbox{$+0.001556 + 0.002046 \simeq +0.0036$ (AU)$^{-1}$} and its orbital
period to a little less than 5000~yr.\\[-0.24cm]

(3c) Accordingly, the motions of the two comets were perturbed during their
Jovian encounters very~unevenly, as their times of closest approach
differed by 165~days, a gap that was the product of their increasing separation
following the parent's earlier fragmentation event.  And while the motion of
C/1973~D1 was~accelerated by the perturbations into an orbit with~a~period of
$\sim$16\,300~yr, the motion of C/1846~O1 was slowed down so profoundly that
this comet ended up in an orbit whose period was shorter by a factor of more
than 3.\\[-0.24cm]

Although the contrast between the orbital periods of the two fragments is
rather astonishing, there are two problems that still remain to be settled.
One is the conditions at the time of the parent's fragmentation needed to
explain the 264-day gap between the two comets, and the second is the process
of shortening the orbital period of C/1846~O1 from the nearly 5000~yr down to
$\sim$1000~yr or less to make the scenario consistent with the range of
periods derived from the observations (Tables~3 and 4).

\subsection{Fragmentation Parameters for the Pair of\\C/1846 O1 and C/1973 D1}
A solution to the first of the two issues is straightforward, because there
exists a precedent:\ the pair of long-period comets C/1988~F1 and C/1988~J1 was
shown in Paper~1 to split off from the parent comet, which was on its way to
perihelion, hundreds of AU from the Sun.  An assumed separation velocity of
about 1~m~s$^{-1}$ in the radial direction was all that was needed to explain
a gap of 76~days between the two comets, measured by the gap between their
times of perihelion passage.


%
The most probable heliocentric distance of the birth of C/1988~F1 and C/1988~J1
was $\sim$400~AU after aphelion, which is equivalent to a time of nearly 700~yr
before the next perihelion.\footnote{The equations that determine the effect of
the fragmentation time and the separation velocity on the perihelion times of
the fragments were provided in Paper~1; they were now applied to C/1846~O1 and
C/1973~D1 as well.}  Given that the orbital period of the 1988 comets was
just about 14\,000~yr (Paper~1), timewise the event took place about 10 times
closer to perihelion than aphelion.

Because the required post-fragmentation gap of 264 days between the modeled
perihelion times of C/1846~O1 and C/1973~D1 was nearly 3.5~times wider and
because the comets were moving in an orbit with a period of 7160~yr far from
the Sun, a separation velocity of 1~m~s$^{-1}$ in the radial direction --- a
representative value based on the arguments presented in Paper~1 --- would
require the fragmentation event to have occurred $\sim$600~AU from the Sun
some 1650~yr before perihelion.  The perturbation effect by Jupiter on
C/1846~O1 was found to increase by a fair amount after we introduced, in
addition, a normal component of the separation velocity of 0.5~m~s$^{-1}$,
since the comet was brought a little closer to the planet at the critical time.
The normal component could not be increased any further in order to keep the
velocity's magnitude from exceeding $\sim$1~m~s$^{-1}$ by a wide margin.  The
details of this phase of our work are in Section~4.

In subsequent runs we tested the dependence of the results on the location of
the fragmentation event in the orbit, at earlier times.  As the required radial
component of the separation velocity was then lower, we could increase the
normal component beyond 0.5 m s$^{-1}$ and further boost the Jovian
perturbations on the comet without violating the total-velocity magnitude
constraint.  For example, moving the fragmentation event to 700~AU from the
Sun along the incoming leg of the parent's orbit, some 2500~yr before
perihelion, required a separation velocity of about 0.4~m~s$^{-1}$ in the
radial direction to accomodate the 264~days gap between C/1836~O1 and C/1973~D1
at perihelion in the 17th millennium BCE, thus allowing us to increase the
velocity's normal component to at least 0.8~m~s$^{-1}$.  Timewise the event
now took place closer to aphelion than to the next perihelion.

Similarly, we investigated a few additional scenarios, in which the
fragmentation event was assumed to have taken place, respectively, at aphelion
and at 700~AU and 500~AU from the Sun before aphelion.  The radial component
required to accommodate the 264-day gap was only a small fraction of
1~m~s$^{-1}$, and the normal component could be increased up to
1~m~s$^{-1}$.  As a special case, we also considered a scenario with the
fragmentation event having occurred shortly after the previous perihelion.
The gap between the two comets at the next perihelion was now increased by
exactly 8 Jupiter's orbital periods (to preserve the required geometry),
allowed by a radial separation velocity slightly exceeding 1~m~s$^{-1}$.

\subsection{Early Post-Fragmentation Motion of C/1846 O1}
%
%
Rigorous orbital computations confirmed both our conclusion (3b) in Section~3.5
concerning the post-encounter orbit of C/1846~O1 and the correlation between
the orbital location of the fragmentation event and the radial component of the
separation velocity.  Consistent with expectation, all 264-day-gap solutions
showed that the comet's initial post-fragmentation perihelion ought to have
taken place in late September through October of $-$14\,326 in order that the
Jovian encounter in December of $-$14\,325 could shorten the orbital period
substantially.

In all scenarios that were run, the first post-encounter orbital period of
C/1846~O1 was below 5000~yr, close to 400 Jovian sidereal periods, and more than
a factor of two shorter than the post-fragmentation, pre-encounter orbital period
of 10\,400~yr at perihelion.  This result is a corollary of the verified conclusion
(3b) of Section 3.5.  The first post-encounter orbital period correlated rather
indistinctly with the normal component of the separation velocity, a higher
velocity leading generally to a slightly shorter period.  The length of the
period did vary to a degree with the orbital location of the fragmentation
event, the shortest period taking place when it occurred near but prior to the
aphelion passage.  This completes the description of the early post-fragmentation
evolution of the modeled motion of C/1846~O1.

\subsection{$\!\!$Stochastic and Encounter-Dominated Perturbations.\\Diffusion
 of Comets}

Consider a comet in an elliptical orbit and unaffected by nongravitational forces.
Let it pass its ``initial'', reference perihelion at time $t_0$ and let its first
return to perihelion occur at time $t_1$, so that the anomalistic orbital period
(in yr),\footnote{Strictly, it is a sidereal orbital period that we need for our
further computations.  In the first approximation, when the anomalistic orbital
period is affected only by the perturbation of the argument{\vspace{-0.04cm}}
of perihelion, $\Delta \omega$ (in deg), the correction (in days) from the
anomalistic to sidereal orbital period equals \mbox{$-0.7174 \, q^{3/2} \Delta
\omega$}.  Since the perturbation $\Delta \omega$ usually amounts to only a small
fraction of 1$^\circ$, the sidereal correction to the anomalistic orbital period
is merely a fraction of 1~day, a change that will be neglected.} which includes
effects of the planetary perturbations integrated over an inteval of time from
$t_0$ to $t_1$, be \mbox{$P_{0,1} = t_1 \!-\!  t_0$}. {\vspace{-0.1cm}}An
effective semimajor axis (in AU) over this time span is \mbox{$a_{0,1} \!=\!
P_{0,1}^{\:\!2/3}$}. {\vspace{-0.15cm}}Let, further, $z_{0,1}$ be the reciprocal
value of $a_{0,1}$, so that \mbox{$P_{0,1} = z_{0,1}^{-3/2}$}.{\vspace{-0.04cm}}
At $t_1$ the comet begins a new revolution about the Sun, during which it is
subjected to the planetary perturbations, whose integrated effect on the
reciprocal semimajor axis is $\Delta z_{1,2}$.  This revolution terminates at
time $t_2$ of the next return to perihelion.  The effective semimajor axis
during this revolution is \mbox{$z_{1,2} = z_{0,1} + \Delta z_{1,2}$}.
Similarly, let $\Delta z_{2,3}$ be an integrated effect of the planetary
perturbations between the perihelion passages at $t_2$ and $t_3$, so that
\mbox{$z_{2,3} = z_{1,2} + \Delta z_{2,3}$}, etc.  For a sequence of perihelion
times, $t_i\;(i \!=\!1,\ldots,n)$, we can write progressively
\begin{eqnarray}
t_1 & \!=\! & t_0 \!+\! z_{0,1}^{-\frac{3}{2}}, \nonumber \\[0.1cm]
t_2 & \!=\! & t_0 \!+\! z_{0,1}^{-\frac{3}{2}} \!\! \left[ 1 \!+\! \left(\!
 1 \!+\! \frac{\Delta z_{1,2}}{z_{0,1}}\!\right)^{\!\!-\frac{3}{2}}
 \right]\!,\nonumber \\[0.1cm]
t_3 & \!=\! & t_0 \!+\! z_{0,1}^{-\frac{3}{2}} \!\! \left[ 1 \!+\! \left(\! 1 \!+\!
 \frac{\Delta z_{1,2}}{z_{0,1}} \! \right)^{\!\!\!-\frac{3}{2}} \!\!\!\!+\! \left(
 \! 1 \!+\! \frac{\Delta z_{1,2}}{z_{0,1}} \!+\! \frac{\Delta z_{2,3}}{z_{0,1}}
 \right)^{\!\!-\frac{3}{2}} \right] \:\!\!\! , \nonumber \\[0.15cm]
  .\,. & \ldots & \ldots\ldots\ldots\ldots\ldots\ldots\ldots\ldots\ldots\ldots
\ldots\ldots\ldots\ldots\ldots\ldots.\,. \nonumber \\[0.15cm]
t_n & \!=\! & t_0 \!+\! z_{0,1}^{-\frac{3}{2}} \!\!\left[ 1 \!+\!\!
 \sum_{k=1}^{n-1} \! \left( \! 1 \!+\!\! \sum_{i=1}^{k} \frac{\Delta
 z_{i,i+1}}{z_{0,1}} \! \right)^{\!\!\!-\frac{3}{2}} \right] \! \ldots
  (n \!\geq\! 2),
\end{eqnarray}
where the time $t_n$ indicates the completion of the comet's $n$-th return to
perihelion.

\begin{figure*} 
\vspace{0.08cm}
\hspace{-0.21cm}
\centerline{
\scalebox{0.893}{
\includegraphics{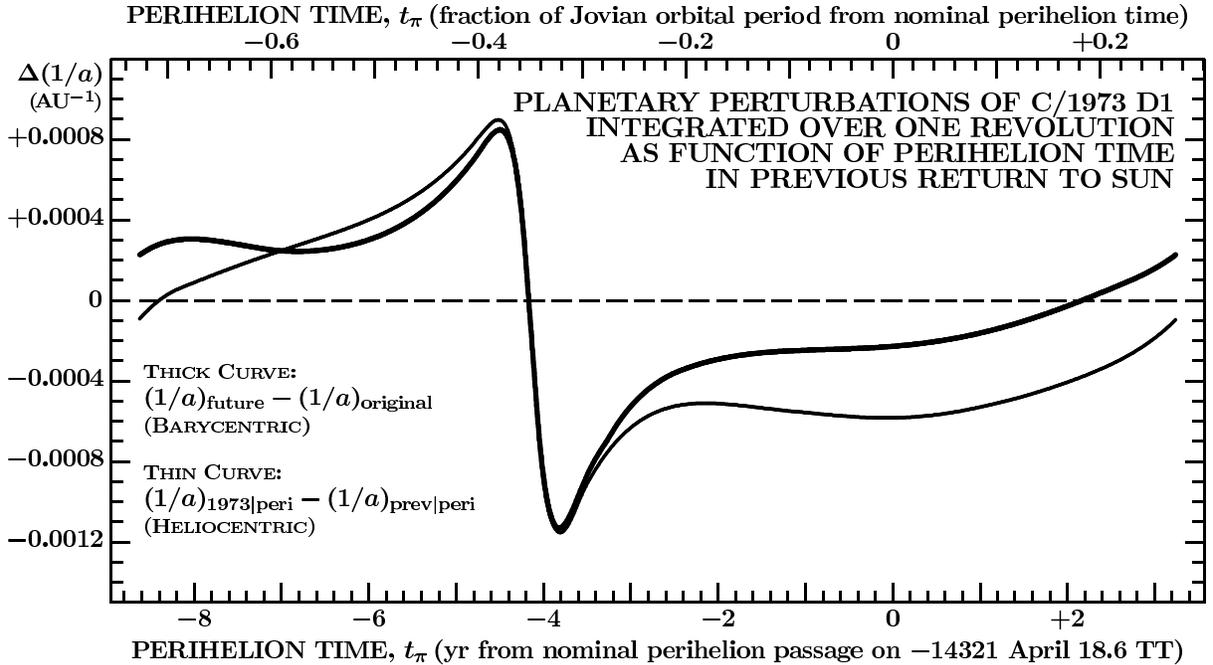}}} 
\caption{Planetary perturbations of the reciprocal semimajor axis of C/1973~D1,
$\Delta(1/a) = \Delta z$, integrated over one revolution about the Sun and
plotted as a function of the perihelion time in the previous return in the 15th
millennium BCE.  The thick curve is a difference in $1/a$ between the ``future''
orbit and the ``original'' orbit at that return in the barycentric system of
coordinates; the thinner curve is a difference between the $1/a$ osculation
values at perihelion in 1973 and at the previous return in the heliocentric
system of coordinates.  In the former case \mbox{$\Delta (1/a) = 0$} marks the
{\vspace{-0.06cm}}future orbit, which is equal to the original orbit relative
to the 1973 return, amounting to +0.0015555 (AU)$^{-1}$.  In the latter case
\mbox{$\Delta (1/a) = 0$} refers to the osculating $1/a$ at the 1973 perihelion,
+0.0009378~(AU)$^{-1}$, and indicates that the orbital period for the
corresponding perihelion time in the 15th millennium BCE was
16\,300~yr.{\vspace{0.5cm}}}
\end{figure*}

In general, the perturbations of the reciprocal semimajor axis of a long-period
comet, integrated over a revolution about the Sun, $\Delta z_{i,i+1}$, follow a
random-walk pattern, being sometimes positive (shortening the orbital period),
sometimes negative (lengthening it).  This process, known as {\it diffusion\/}
of comets, may eventually lead to a comet's capture into a tightly-bound orbit
or to its expulsion from the Solar System into interstellar space.

A stochastic evolution of cometary orbits driven by the planetary perturbations
was many times investigated in connection with a steady-state distribution of
observed comets and with the hypotheses of comet origin.  Following a pioneering
work by van Woerkom (1948), an enormous progress was recently achieved in the
understanding of the diffusion process of objects from the Oort Cloud, Kuiper
Belt, and Scattered Disk by applying powerful Monte Carlo numerical-simulation
techniques and methods of long-term numerical integration.  Most of these
advances were summarized and reviewed by Dones et al.\ (2004), Morbidelli \&
Brown (2004), Duncan et al.\ (2004), Rickman (2004), and others.

Figure 4 displays the planetary perturbations of the reciprocal semimajor axis,
integrated over one revolution about the Sun, as a function of the perihelion
time, $t_\pi$, of C/1973~D1 at its previous return in the 15th millennium BCE;
both the standard difference between the barycentric ``future'' and ``original''
orbits (thick curve) and the difference between the heliocentric osculating
values at perihelion in 1973 and at the previous return (thin curve) are
presented.  The two curves have somewhat similar features, but are by no means
alike.  The most remarkable feature is in a relatively narrow range of perihelion
times centered on $-$4.12~yr, which allows the comet to have the closest possible
approach to Jupiter, to 0.649~AU.  In this range of $t_\pi$ the two curves copy
each other very closely, the only perceptible difference being in the amplitude:\
0.002046~(AU)$^{-1}$ in the heliocentric system, but 0.001969~(AU)$^{-1}$ in the
barycentric system.  

Figure~4 further shows that, statistically, the pattern of perturbations,
dominated by the encounter with Jupiter, slightly favors outward diffusion,
toward larger orbits and longer orbital periods.  This tendency is numerically
demonstrated by computing an integrated perturbation, \mbox{$(1/a)_{\rm future}
\!-\!  (1/a)_{\rm original}$}, averaged over the Jovian orbital period,
$P_{\rm J}$: 
\begin{equation}
\langle \Delta (1/a)_{\rm orig,fut} \rangle = \langle \Delta z \rangle =
 \frac{1}{P_{\rm J}} \int_{t_{\rm beg}}^{t_{\rm beg} \!+\!  P_{\rm J}}
 \!\!\!\Delta z(t_\pi) \, dt_\pi,
\end{equation}
where $t_{\rm beg}$ is a chosen integration start time.  Integrating over the
thick curve in Figure~4, we find that \mbox{$\langle \Delta z \rangle =
-0.0000334$ (AU)$^{-1}$}.

Our scenario for C/1846 O1, charted in Section 3.5,~is statistically atypical.
To estimate a probability and magntidue time scale over which C/1846~O1 would
under random diffusion conditions evolve from its initial post-encounter orbit
into an orbit of a much shorter period, comparable to the orbit observed in
1846, we integrated the portion of the thick curve in Figure~4 over which
\mbox{$\Delta z(t_\pi) > 0$} to find \mbox{$\langle \Delta z^\ast \rangle =
+0.000322$ (AU)$^{-1}$}.  The range of these perihelion times equals 5.4862~yr
or 0.463~the Jovian orbital period.  Thus, the probability of a comet in the
orbit of C/1973~D1 having its period shortened during one perihelion passage
is 0.463; after $n$ returns, the probability is 0.463$^n$.  To estimate $n$,
we recall from Section 3.5 that at the first return to perihelion
\mbox{$z_{0,1} = z_{\rm init} \simeq +0.0036$ (AU)$^{-1}$} and replace
\mbox{$\Delta z_{i,i+1}\;(i \!=\!  1,\,2,\ldots)$} in Equations (3) with
$\langle \Delta z^\ast \rangle$ to simplify the expression; for \mbox{$n
\geq 2$} we obtain:
\begin{eqnarray}
t_n & = & t_0 + P_{\rm init} \!\left[ 1 +\!  \sum_{k=1}^{n-1} \left(\! 1 + k \,
 \frac{\langle \Delta z^\ast \rangle}{z_{\rm init}} \right)^{\!\!-\frac{3}{2}}
 \right] \nonumber \\[0.1cm]
    & = & t_0 + P_{\rm init} \sum_{k=0}^{n-1} \left( \! 1 + k \, \frac{\langle
 \Delta z^\ast \rangle}{z_{\rm init}} \right)^{\!-\frac{3}{2}} \!\!,
\end{eqnarray}
where \mbox{$P_{\rm init} = P_{0,1} = t_1 - t_0 = z_{\rm init}^{\frac{3}{2}}
\approx 4630$ yr}.  According to Figure 3, the proposed scenario requires the
initial perihelion time, $t_0$, in late September through October of the year
$-$14\,326.  The time scale, over which the comet's orbital period should get
reduced below a required limit, $P_{\rm lim}$, equals the interval \mbox{$t_n
- t_0$}.  The
%
%
%
%
%
final orbital period in 1846, \mbox{$P_{\rm fin} = P_{n-1,n}$}, is related to
the initial period by
\begin{equation}
P_{\rm fin} = P_{\rm init} \! \left[ 1 + (n-1) \, \frac{\langle \Delta z^\ast
\rangle}{z_{\rm init}} \right]^{-\frac{3}{2}} \!\! ,
\end{equation}
and since \mbox{$P_{\rm fin} < P_{\rm lim}$},
\begin{equation}
n > 1 + \frac{z_{\rm init}}{\langle \Delta z^\ast \rangle} \! \left[ \left( \!
\frac{P_{\rm init}}{P_{\rm lim}} \! \right)^{\frac{2}{3}} \!\!- 1\right] \!.
\end{equation}
%
%
%
%
%
%
Requiring, for example, that \mbox{$P_{\rm lim} \approx 1000$ yr}, the above
values of $\langle \Delta z^\ast \rangle$, $z_{\rm init}$, and $P_{\rm init}$
imply that \mbox{$n > 20$}.  With \mbox{$n = 21$} the time scale \mbox{$t_n-t_0
= 9.59 \, P_{\rm init} \simeq 44\,400$ yr}, nearly three times longer than the
case of a genetical relationship would require for C/1846~O1.  The probability
of this systematic reduction of the orbital period to happen under random
diffusion conditions is \mbox{0.463$^{21} \simeq 10^{-7}$}.  One could use
Figure~4 to come up with other similar scenarios with even lower probability.

It is invariably assumed that the diffusion process is governed by the Gaussian
law.  Zhou et al.\ (2002) pointed out, however, that the Gaussian approximation
is appropriate only for small perturbations of $1/a$ accumulating over a number
of revolutions.  When an average perturbation per revolution is not small in
comparison with the final change in $1/a$, Zhou et al.\ proposed that the
orbital evolution is governed by the L\'evy (1937) random walk because of a
disproportionately large contribution by the significant perturbation events due
to close approaches to Jupiter.  The L\'evy walk appears to better fit enhanced
(or anomalous) diffusion and has a number of applications in physics (e.g.,
Shlesinger et al.\ 1987).

The paradigm of random walk, applicable to a statistical sample, has no
prognosticative merit in an individual case.  In fact, the best known
triggers of dramatically enhanced rates of diffusion in the orbital
evolution of comets are the occasional very close encounters with Jupiter
(\mbox{$\Delta_{\rm J} < 0.1$ AU}), during which an orbit can be transformed
beyond a shade of recognition.  They happen in spite of their extremely low
{\it a priori\/} probabilities of occurrence.  An example is C/1770~L1
(Lexell), observed at a single apparition as a short-period~comet~with a
period of 5.60~yr and approaching the Earth to a record minimum distance
of 0.0151~AU on 1770 July~1 (Sekanina~\& Yeomans 1984).  The comet's orbital
history was investigated more than once (e.g., Lexell 1778; Le Verrier 1857;
Callandreau 1892; Kazimirchak-Polonskaya 1967, 1972; Carusi et al.\ 1985).
There is a general consensus that the comet had very close encounters with
Jupiter shortly before discovery and again a dozen years later.  The results of
the computations by Kazimirchak-Polonskaya (1967) show that the first encounter
took place in March 1767 (\mbox{$\Delta_{\rm J} = 0.020$ AU}) and the second
one in July 1779 (\mbox{$\Delta_{\rm J} = 0.0015$ AU}).  The total perturbation
{\vspace{-0.04cm}}effect of the first event was about \mbox{$\Delta z = +0.106$
(AU)$^{-1}$}, of{\vspace{-0.04cm}} the second event at least \mbox{$\Delta z =
-0.32$ (AU)$^{-1}$}.  These energy jumps are two orders of magnitude greater
than the peak integrated perturbation effect for C/1973~D1 in Figure~4.

\subsection{High-Order Orbital-Cascade Resonance}
Isolated close encounters with Jupiter, ruled out in the case of C/1846~O1, are
not the only means to distinctly disrupt a comet's evolution of slow, random
orbital diffusion.  Another well-known example of a strongly nonradom pattern
of cometary motions is orbital (mean-motion) resonance, defined classically as
a periodic gravitational influence of a perturbing planet (usually Jupiter) due
to the two bodies' orbital periods having a ratio of two small integers.  The
result can be either a destabilization of the comet's orbit or a pattern of a
recurring configuration (libration) over a certain period of time, depending
on whether relatively close encounters keep occurring or are systematically
avoided and on how close to being perfect the resonance is.

We already pointed out in Section 3.7 that in our early computer runs the
initial post-fragmentation orbital period for C/1846~O1 --- near 400~Jovian
sidereal periods following the comet's approach to the planet some 440~days
after perihelion (for details, see Section~4) --- was followed by a still
shorter period by the next return.  Since this particular timing of the
encounter was instrumental in a significant reduction of the orbital period,
a repetitive, long-term trend of this kind should warrant a scenario in which
the comet kept encountering Jupiter $\sim$440~days after perihelion at as many
consecutive returns as possible.  And since, on the other hand, a continuing
recurrence of this configuration required that the comet's perihelion passages
followed each other after an integral number of Jovian revolutions about the
Sun, this replicate mean-motion commensurability --- capable to systematically
reduce the orbital period of C/1846~O1 quite dramatically over a relatively
short period of time --- locked the comet's motion temporarily in what we
refer to as a {\it high-order orbital-cascade resonance\/} with Jupiter's
orbital motion. An {\it a priori\/} probability of this lock is low, but so
is the probability of close encounters with Jupiter (Sec.~6).

The term high-order orbital-cascade resonance requires an explanation, because
two fundamental properties of orbital resonance, as usually understood in
celestial mechanics, are missing.  One, the comet's and Jovian orbital periods
are {\it not\/} related by a ratio of two {\it small\/} integers (whence {\it
high-order\/}) and the integers decrease from return to next return; and,
two, the result of the resonance is neither a stabilization (libration) nor
a destabilization of the orbit, but its period's rapid and profound
shortening that proceeds in successive discrete steps (whence {\it cascade\/})
over a span of time that is only a factor of two or so longer than the
pre-fragmentation orbital period. 

The repetition, from return to next return, of nearly identical encounter geometry
is demanded in order to preserve a recurrence of nearly constant integrated
perturbations, $\Delta z_{\rm res}$.  Athough our reason is now different, we
nonetheless require a formal modification of Equations~(3) that is the same as
in Equations~(5),
\begin{equation}
t_n = t_0 + P_{\rm init} \sum_{k=0}^{n-1} \left( \! 1 + k \, \frac{\Delta
 z_{\rm res}}{z_{\rm init}} \! \right)^{\!\!-\frac{3}{2}} \!\!,
\end{equation}
except that we solve this equation for $\Delta z_{\rm res}$ rather than for $n$
and, consequently, there are multiple solutions that depend on the choice of
$n$.  In addition, since we try to fit the initial conditions related to the
arrival of C/1846~O1, $t_n$ is the comet's observed perihelion time, \mbox{$t_n
= 1846.40$}, and therefore a constraint rather than an unknown.  Equation~(8)
has no solution for \mbox{$n < n_0$}, where $n_0$ is a minimum number of returns
needed in order that \mbox{$\Delta z_{\rm res} > 0$}.  Introduction of another
condition, \mbox{$P_{\rm fin} < P_{\rm lim}$} (analogous to that used in
Section~3.8), may restrict the number of returns more severely, to \mbox{$n >
n_{\rm lim}$}.

Calling \mbox{$\Delta \zeta = \Delta z_{\rm res}/z_{\rm init}$} and
isolating the known parameters $t_0$, $t_n$, and $P_{\rm init}$ on one
side, we rewrite Equation~(8) in terms of dimensionless quantities $\Re$,
$\Delta \zeta$:
\begin{equation}
\Re = \frac{t_n \!-\! t_0}{P_{\rm init}} = \! \sum_{k=0}^{n-1} (1 +
 k\,{\Delta \zeta} \:\! )^{-\frac{3}{2}} ,
\end{equation}
where $\Re$ is an allowed normalized orbit-evolution time.  The solutions
\mbox{$\Delta \zeta = f(\Re,n)$} were derived by a method of successive
approximations for a wide range of $n$, the number of returns to perihelion.
First, however, we needed information from orbit-integration runs aimed at
the evolutionary scenario proposed in Section~3.5.

\begin{table}[b] 
\vspace{-3.3cm}
\hspace{4.23cm}
\centerline{
\scalebox{1}{
\includegraphics{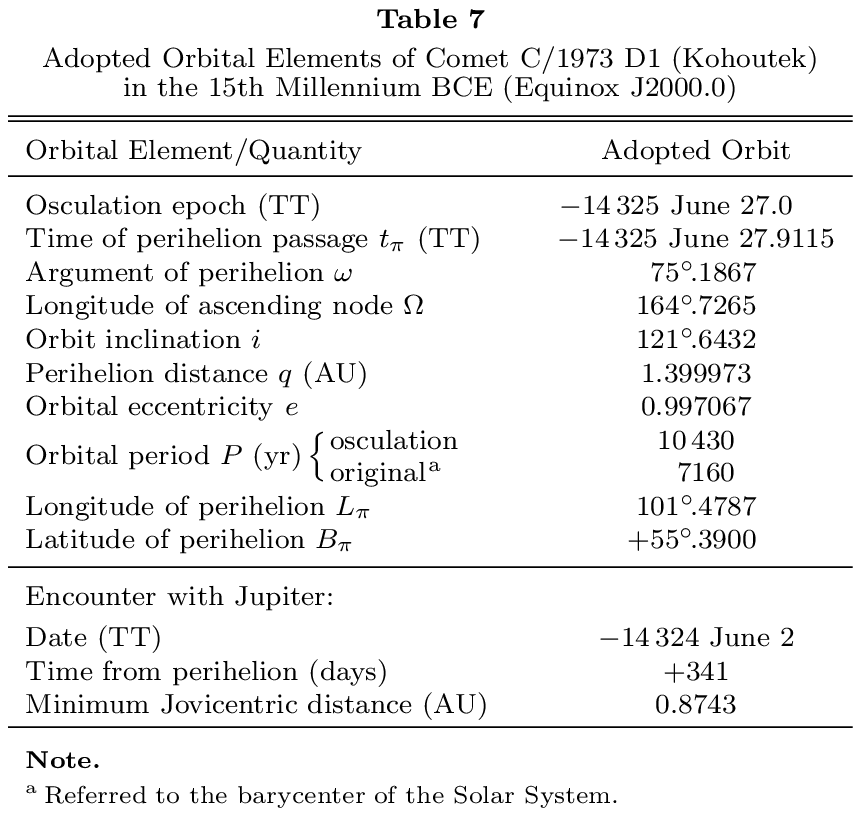}}} 
\vspace{-17.95cm}
\end{table}

\section{Early Orbit-Integration Runs}

In compliance with the proposed scenario, the set of elements for C/1973~D1
with a minimum osculating orbital period of 10\,430~yr at perihelion, on
$-$14\,325~June~27, was chosen as the starting orbit (Table 7), whose formal
errors are those in Table~4.  Its choice, as a product of integration of
the comet's 1973 orbit back in time with an extremely slight correction
to the nominal value of the eccentricity, assures us that the 1973 orbit
of C/1973~D1 should satisfy any fragmentation scenario, provided that
the separation velocity is added to C/1846~O1.  Since C/1973~D1 was a
companion (Section~2), the presumed parent of C/1846~O1 and C/1973~D1 had
been moving in an orbit of slightly shorter period than C/1973~D1 and
required that a separation velocity of the primary be formally referred to
the companion rather than the other way around, as is customary when
determining the conditions at fragmentation.

In compliance with the proposed scenario, the next step was the choice of the
fragmentation time that would assure C/1846~O1 to arrive at perihelion about
264~days before C/1973~D1, that is,{\vspace{-0.02cm}} around $-$14\,326 October
6; and, more importantly, to accomplish a close approach to Jupiter 165~days
before C/1973~D1, that is, on or around $-$14\,325 December 20.  In the early
runs, we searched for a solution that would satisfy the premise of a separation
velocity between C/1973~D1 and C/1846~O1 of 1~m~s$^{-1}$ in the radial
direction (C/1846~O1 sunward), because this component has by far the
most significant effect on the subsequent perihelion time.  The first solution
we tested was for a fragmentation event at a heliocentric distance of 569~AU,
about 1470~yr before perihelion, that is, in approximately 15\,800~BCE.  This
solution satisfied the perihelion-time constraint to within two weeks, but
fitted the Jovian-encounter condition very well.

\begin{table*}[t] 
\vspace{-3.85cm}
\hspace{-0.5cm}
\centerline{
\scalebox{1}{
\includegraphics{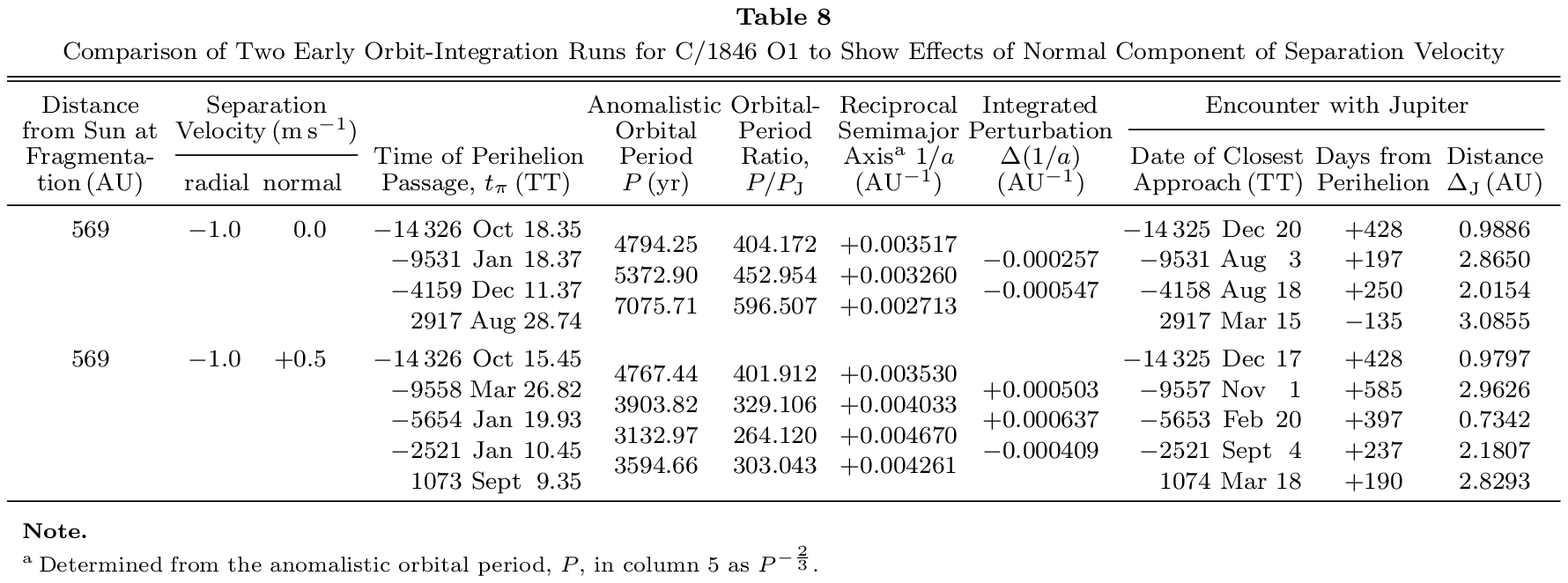}}} 
\vspace{-18.7cm}
\end{table*}

The results of integration are presented in the upper half of Table~8.  While,
as expected [cf.\ condition (3b) in Section~3.5], the comet returned to
perihelion in less than 5000~yr, the successive orbital periods grew
progressively longer, contrary to our expectation.  This solution was
obviously unacceptable.

Since Figure 2 shows that the semimajor axis is most perturbed outside the
node (that is, when Jupiter is out of the comet's orbit plane), we considered
it desirable to introduce, in addition, an out-of-plane component of
the separation velocity.  Rather arbitrarily, we chose 0.5~m~s$^{-1}$
and found that the direction above the plane is the one that decreases
the encounter's miss distance.  Integration of this solution --- the
small normal component of the separation velocity being the {\it only\/}
difference --- offered dramatically different results, as shown in the
lower half of Table~8.  While the effects on the initial perihelion time
and the Jovian-encounter time are only 3 days and on the initial orbital
period merely 27~yr (less than 0.6\%), the times of the next return to
perihelion already differed by $\sim$1500~yr, a major trend that continued.
By the year 1073, the comet completed four revolutions about the Sun in
the second orbit, but less than three revolutions when in the first orbit.

The enormous discrepancy in the orbital motion of the comet in the two tested
orbits was caused not only by the slightly smaller encounter distance, but
primarily by the fact that the encounter times in the first three passages
were --- contrary to the random-walk rule --- always on the ``correct'' side
of the node, $\sim$400~days or more after perihelion, which warranted that
the integrated perturbation $\Delta(1/a)$ was positive, as Table~8 plainly
indicates.  Only in the fourth return the closest approach to Jupiter occurred
less than 300~days after perihelion and the gradual shortening of the orbital
period came to an end.

\begin{figure}[b] 
\vspace{0.6cm}
\hspace{-0.18cm}
\centerline{
\scalebox{0.715}{
\includegraphics{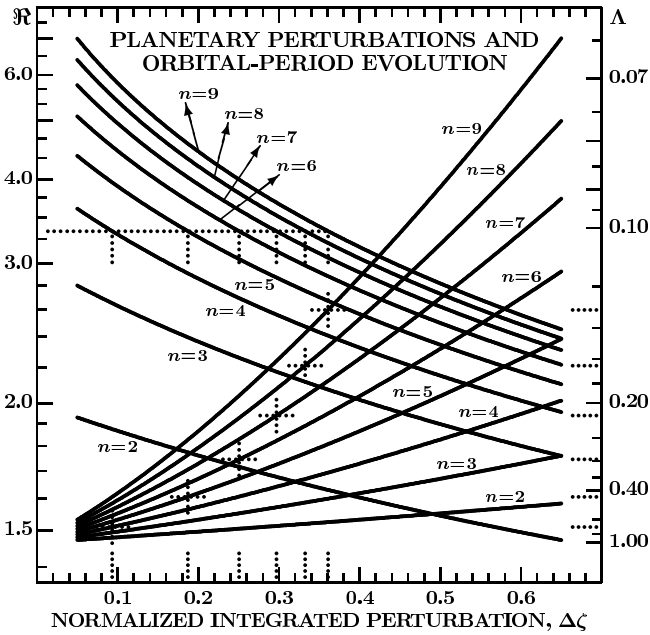}}} 
\caption{Plot of an allowed normalized orbit-evolution time, $\Re$ (the curves
that decrease from the upper left to the lower right, with the scale on the left),
and the final orbital period as a fraction of the initial orbital period,
\mbox{$\Lambda = P_{\rm fin}/P_{\rm init}$} (the curves that increase from the
lower left to the upper right, with the scale on the right), against a resonance
perturbation of the reciprocal semimajor axis integrated over one revolution
about the Sun in units of the initial reciprocal semimajor axis, $\Delta \zeta$.
Keeping the number of revolutions about the Sun, $n$, to less than 10, the dotted
lines show that, for \mbox{$\Re = 3.392$}, the allowed solutions (left scale) are
\mbox{$\Delta \zeta = 0.0934$}, 0.1871, 0.2505, 0.2968, 0.3325, and 0.3609 for,
respectively, \mbox{$n = 4$}, 5, 6, 7, 8, and 9, which with \mbox{$z_{\rm init}
= 0.003530$}~(AU)$^{-1}$ correspond to \mbox{$\Delta z_{\rm res} = 0.00033$},
0.00066, 0.00088, 0.00105, 0.00117, and 0.00130~(AU)$^{-1}$.  There are no
solutions for \mbox{$n < 4$}.  The solutions for $P_{\rm fin}/P_{\rm init}$
are 0.6906 with \mbox{$n = 4$}, 0.4326 with \mbox{$n = 5$}, 0.2958 with
\mbox{$n = 6$}, 0.2156 with \mbox{$n = 7$}, 0.1648 with \mbox{$n = 8$}, and
0.1305 with \mbox{$n = 9$}.  With \mbox{$P_{\rm init} = 4767.44$}~yr, the final
orbital periods $P_{\rm fin}$ are, respectively, 3292~yr, 2062~yr, 1410~yr,
1028~yr, 786~yr, and 622~yr.  A constraint of, for example, \mbox{$P_{\rm fin}
< 1600$}~yr allows only $n$ equal to 6, 7, 8 or 9.  There are also constraints
on $\Delta z_{\rm res}$ (see text).{\vspace{0.07cm}}}
\end{figure}

Another useful purpose served by this exercise was a determination of the
magnitudes of the integrated perturbation of the semimajor axis in these
scenarios.  For the positive changes of $\Delta (1/a)$ Table~8 suggests a
range{\vspace{-0.05cm}} of 0.0005 to 0.0006~(AU)$^{-1}$, which is close
{\vspace{-0.06cm}}to an average (e.g., van Woerkom 1948; \v{S}teins \&
Kronkalne 1964; Fern\'andez \& Gallardo 1994).

A practical impact of this result on our further investigation had to do with
the solution to Equation~(9), as the unknown $\Delta \zeta$ depends critically
on an integrated perturbation of the reciprocal semimajor axis in the case of
orbital-cascade resonance.  In Figure~5 we plot the allowed normalized
orbit-evolution time $\Re$ and the ratio \mbox{$\Lambda = P_{\rm fin}/P_{\rm
init}$} as a function of the normalized perturbation $\Delta \zeta$.  To
determine $\Re$, we use \mbox{$t_n = 1846.40$} for any $n$ and employ the
constants from the second scenario in Table~8, \mbox{$t_0 = -14\,325.21$ yr} and
\mbox{$P_{\rm init} = 4767.44$ yr}; we find \mbox{$\Re = 16\,171.61/4767.44 =
3.392$}, which gives us six different solutions for \mbox{$n = 4, \ldots ,9$}.
We read the values of $\Delta \zeta$ on the axis of abscissae and compute the
integrated resonance perturbation \mbox{$\Delta z_{\rm res} = 0.00353 \, \Delta
\zeta$ (AU)$^{-1}$}.  We also read the ratio \mbox{$\Lambda = P_{\rm fin}/P_{\rm
init}$} on the right-hand side of the axis of ordinates and compute the final
orbital period, \mbox{$P_{\rm fin} = 4767.44 \, \Lambda$ yr}.  The caption to
Figure~5 describes the individual solutions in detail.  The end result is that
the final period \mbox{$P_{\rm fin} < 1600$ yr} allows only \mbox{$n \ge 6$},
whereas \mbox{$\Delta z_{\rm res} < 0.0007$ (AU)$^{-1}$} (cf.\ Table~8) allows
only \mbox{$n = 4$ or 5}; the two conditions are not satisfied simultaneously
for any $n$.

At this point of our experimentation, the conclusion was that if C/1846~O1 and
C/1973~D1 are genetically related, then, in the least, it is dynamically extremely
unlikely in the presence of orbital-cascade resonance that a fragmentation event
occurring at $\sim$569~AU before a perihelion passage in the 15th millennium BCE
could ``launch'' C/1846~O1 into an orbit that should eventually (in the 19th
century) have an orbital period much shorter than about 2000~yr.  We show in
Section~5 that, fortunately, the motion of C/1846~O1 could be subjected to
greater integrated perturbations of the semimajor axis, if the fragmentation
event took place earlier, nearer the aphelion of the parent orbit.  That option
has another significant advantage:\ it needs a lower radial component of
the separation velocity to keep C/1846~O1 and C/1973~D1 apart at the required
$\sim$264~days at perihelion.  As a result, it is possible to increase the
magnitude of the normal component and still hold the total separation velocity
at a realistic level near 1~m~s$^{-1}$.

In deriving the two solutions presented in Table 8, no attempt was made to
bring the orbital periods closer to a commensurability with the orbital period
of Jupiter.  Yet, the second solution shows that, of the four periods listed,
the best is commensurable within 0.043 and the worst within 0.120 the Jovian
period.  The same computer code that was used to generate Figure~5 was also
employed to investigate the commensurability and thus the chances of high-order
orbital-cascade resonance to more significantly affect the rate of the
orbital-period's reduction, as was discussed in Section~3.9. 

\begin{table*}[ht] 
\vspace{-4.15cm}
\hspace{-0.535cm}
\centerline{
\scalebox{1}{
\includegraphics{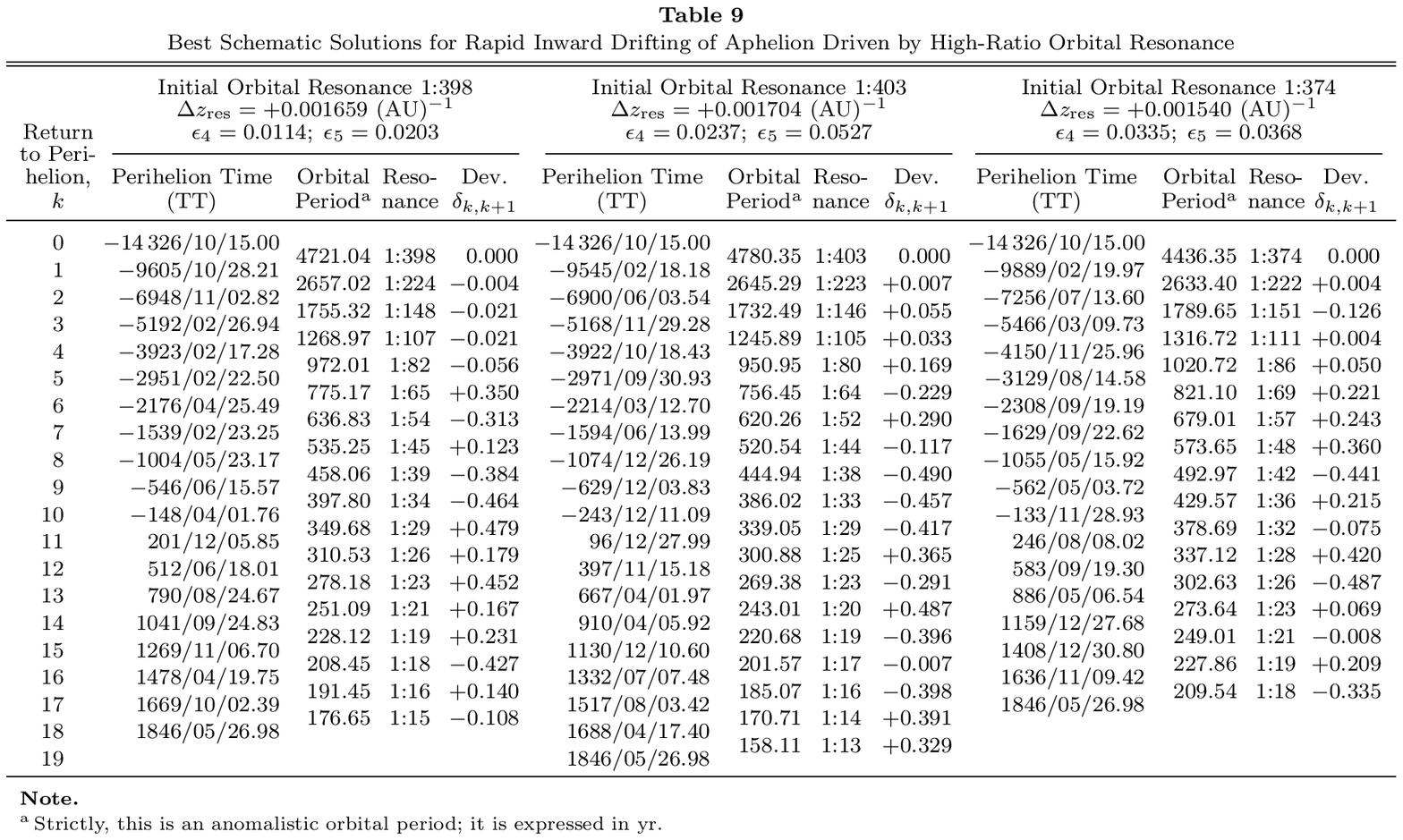}}} 
\vspace{-14.25cm}
\end{table*}

\section{REFINED ORBIT-INTEGRATION RUNS}
For a given heliocentric distance at fragmentation, the timing of the comet's
subsequent arrival to perihelion could always be adjusted by slightly varying
the separation velocity's radial component, so that the initial orbital period,
$P_{\rm init}$, be minimized.  A very minor additional adjustment would be
needed to assure that the period $P_{\rm init}$ be, in addition, nearly
perfectly commensurable with the Jovian period.
%

It was this experimentation that prompted us to investigate, by trial and
error, the most consistent sequences of orbital periods as a function of
the allowed orbital-evolution time, $\Re$, and the number of returns, $n$,
with $\Re$ from Equation~(9) now expressed in terms of the 1:$\Gamma_{0.1}$
commensurability of the initial orbital period, \mbox{$P_{0,1} = P_{\rm
init}$} with the Jovian period, $P_{\rm J}$:
\begin{equation}
P_{\rm init} = \Gamma_{0,1}P_{\rm J}.
\end{equation}
where $\Gamma_{0,1}$ is an integer.

The nature of changes in the orbital period --- especially in cases considered
here, when a perturbation of the reciprocal semimajor axis integrated over
a revolution about the Sun, $\Delta z_{\rm res}$, is essentially constant
from orbit to orbit --- is such that the period drops most substantially
over the first several returns after the fragmentation event.  These early
returns are accordingly the primary target of our interest, so much so in
fact that they are the only ones over which the orbital period needs to be
locked in a temporary cascade resonance to make the period $P_{\rm fin}$
broadly consistent with the observations.  The degree of solutions' compliance
with such an evolution of the orbital period could be tested by the computer
code via an averaged deviation of the first $m$ periods from strict
commensurablity:
\begin{equation}
\epsilon_m = \frac{1}{m} \sum_{i=1}^{m} \min [{\sf mod}\langle \chi_{i-1,i},
1 \rangle; 1 \!- {\sf mod}\langle \chi_{i-1,i}, 1 \rangle],
\end{equation}
where
\begin{equation}
\chi_{i-1,i} = \frac{P_{i-1,i}}{P_{\rm J}},
\end{equation}
{\sf mod} is the {\it modulo operation\/}'s remainder,\footnote{Typically,
the {\it modulo operation\/} is described symbolically as \mbox{$\alpha \;
{\sf mod} \; k$} or \mbox{$\alpha \; ({\sf mod} \; k)$}, where $\alpha$
(here a floating-point quantity) is a dividend, integer $k$ is a divisor,
${\sf mod}\langle \alpha, k\rangle$ is a remainder, and integer
\mbox{$Q = \alpha/k \!-\!  {\sf mod}\langle \alpha, k \rangle$} is a
quotient.} and $m$ equals 4 or 5.  We already said that the initial orbital
period can always be made perfectly consistent with an appropriately chosen
commensurability 1:$\Gamma_{0,1}$; the nearest commensurabilities of the
successive orbital periods, 1:$\Gamma_{1,2}$, 1:$\Gamma_{2,3}$, etc., are
\begin{equation}
\Gamma_{i-1,i} = \chi_{i-1,i} - {\sf mod}\langle \chi_{i-1,i},1 \rangle,
 \;\;\; (i = 2, 3, \ldots),
\end{equation}
when \mbox{${\sf mod}\langle \chi_{i-1,i},1 \rangle < \frac{1}{2}$}, but
\begin{equation}
\Gamma_{i-1,i} = 1 + \chi_{i-1,i} - {\sf mod}\langle \chi_{i-1,i},1 \rangle,
 \;\;\; (i = 2, 3, \ldots),
\end{equation}
when \mbox{${\sf mod}\langle \chi_{i-1,i},1 \rangle > \frac{1}{2}$}.  Their
{\vspace{-0.04cm}}averaged deviation $\epsilon_m$ from strict cascade resonance
is given by Equation~(11).

\begin{table*}[ht] 
\vspace{-4.1cm}
\hspace{-0.535cm}
\centerline{
\scalebox{1}{
\includegraphics{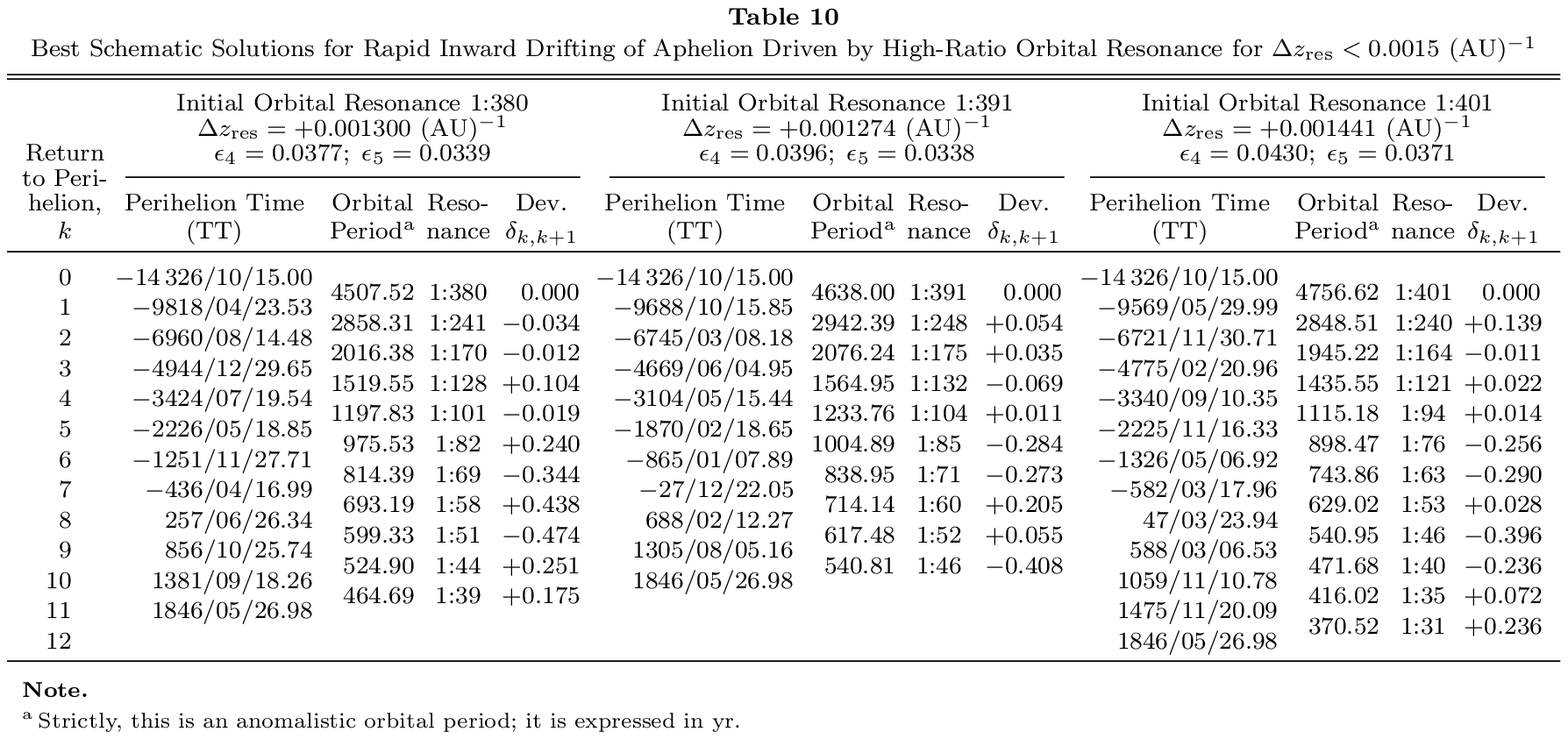}}} 
\vspace{-16.6cm}
\end{table*}

For \mbox{$370 \leq \Gamma_{0,1} \leq 430$}, three among the orbital-period
sequences that were found to be most consistent with cascade resonance (lowest
values of $\epsilon_m$) are {\nopagebreak}presented in Table~9; their purpose
was to guide us in our effort to come up with solutions that best mimicked the
orbital period of C/1846~O1, estimated from the observations.  For each sequence
the numbers in the column ``Resonance'' are derived from Equation~(13) or (14),
and the individual deviations from the exact resonance, in the last column, from
\begin{equation}
\delta_{k,k+1} = \chi_{k,k+1} - \Gamma_{k,k+1}, \;\;\;(k = 0,1, \ldots, n\!-\!1),
\end{equation}
where it always holds that \mbox{$\delta_{0,1} = 0$} because of the condition
(10).  We note that, by a fifth return to perihelion, the orbital period drops
to near or below 1000~yr, in fair agreement with the period of C/1846~O1 derived
from the observations in Section~3.2.  Accordingly, in cases like these we do
not need to have cascade resonance locked for longer than five returns to the
Sun, with random walk variations from a sixth return on making little difference
except for the condition of the perihelion date in 1846.  Indeed, Table 9 shows
that a presumed continuation of the pattern of systematic reduction after the
fifth revolution would lead to an orbital period of \mbox{150--200}~yr in 1846,
far shorter than dictated by the observations.

It would seem that because of our interest in the conditions at only the
several early returns to the Sun, the total number, $n$, of chosen returns
is irrelevant.  This unfortunately is not so, because the integrated
perturbation $\Delta z_{\rm res}$ increases with $n$.  The three solutions
in Table~9, whose $n$ is, respectively, 18, 19, and 17, were selected from
a total of about 800 solutions with \mbox{$n \leq 20$}.  Their perturbation
$\Delta z_{\rm res}$ always exceeds +0.0015~(AU)$^{-1}$ and is about three
times as high as the relevant integrated perturbations $\Delta (1/a)$ in
Table~8.


Judging from the unrealistically short orbital periods $P_{\rm fin}$, the
solutions in Table~9 appear to be too powerful.  To correct this problem and
simultaneously reduce the magnitude of the integrated perturbation $\Delta
z_{\rm res}$, we next restricted our search only to solutions with
\mbox{$n \leq 12$}, which imply generally \mbox{$\Delta z_{\rm res} < +0.0015$
(AU)$^{-1}$}.  Selected from a total of about 300 solutions, the three that
fit best the conditions of cascade resonance are presented in Table~10, whose
format is identical with that of Table~9.

An interesting, but apparently entirely fortuitous property of each of the
three solutions in Table~10 is that an averaged deviation from a strict
commensurability of the first five orbital periods, $\epsilon_5$, is lower
than that of the first four ones, $\epsilon_4$.  This indicates that the
fifth period fits an integral multiple of the Jovian orbital period better
than the average of the first four.  In Table~9, the opposite was true in
each case.  By the time of the sixth return, when the resonance lock has
been lost, the comet's orbital period is already near 1000~yr, so in
this sense the solutions in Table~10 are almost as openly disposed to
resonance as those in Table~9.  The implied perturbation $\Delta z_{\rm
res}$ in Table~10 is by about 300 units of 10$^{-6}$\,(AU)$^{-1}$ lower
compared to Table~9, which is still a little more than twice as high as
the relevant $\Delta (1/a)$ in Table~8.

\begin{table*}[ht] 
\vspace{-3.8cm}
\hspace{-0.52cm}
\centerline{
\scalebox{1}{
\includegraphics{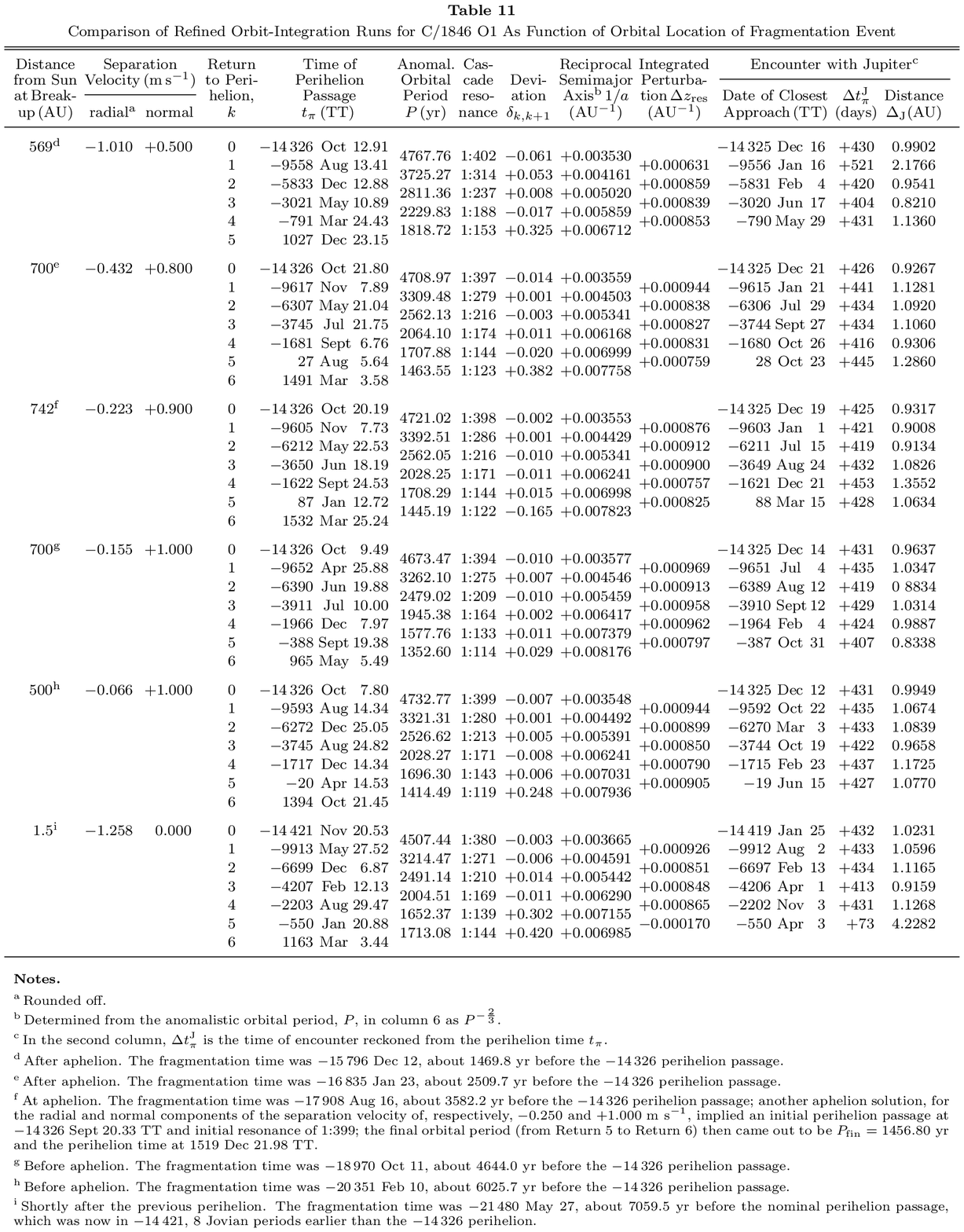}}} 
\vspace{-2cm}
\end{table*}

In the context of comparing these universal schemes with the actual orbit
integrations of the modeled motion for C/1846~O1, it should be recognized
that the schemes serve only to guide us toward assessing our chances for
achieving successful solutions to the problem of rapid systematic inward
drifting of aphelion by orbital-cascade resonance.  The schemes ignore the
dependence on the heliocentric distance at fragmentation, disregard effects
of the indirect planetary perturbations (unrelated to the Jovian encounter
some 440~days after perihelion), and require that the integrated
perturbation $\Delta z_{\rm res}$ be strictly invariable from orbit to
orbit.  In practical integrations, these conditions are of course not
satisfied, so that the schemes are of only limited assistance.

\begin{table*}[t] 
\vspace{-3.85cm}
\hspace{-0.52cm}
\centerline{
\scalebox{1}{
\includegraphics{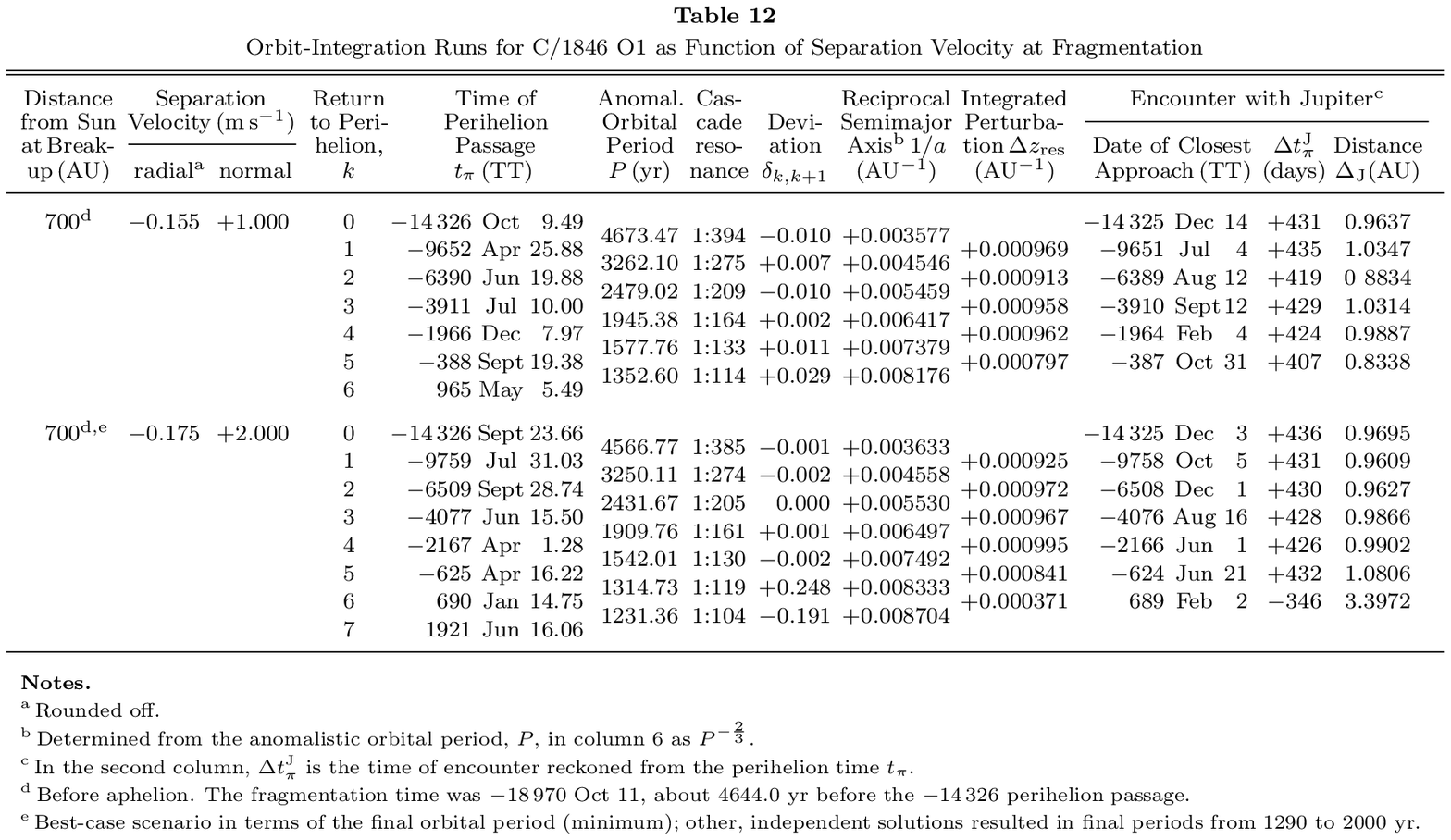}}} 
\vspace{-15cm}
\end{table*}

Nonetheless, the schemes were quite helpful in demonstrating that near-perfect
cascade resonance conditions, produced by the constant integrated perturbations
of the reciprocal semimajor axis, are compatible with a broader random-walk
pattern.  They can in fact extend over at least four to five consecutive
revolutions about the Sun, during which the rate of systematic inward drifting
of aphelion is high enough to lead, at the time the resonance lock brakes down,
to an orbital period that is about equal to, or shorter than, 1000~yr and thus
comparable to the orbital period of C/1846~O1 dictated by the observations
(Section~3.2).  Our experimentation confirmed that the integrated
perturbation effect $\Delta z_{\rm res}$ was a function of the location of
the fragmentation event in the parent comet's orbit, so that the significance of
the apparent discordance between the values of $\Delta z_{\rm res}$ in Table~8
on the one hand and Tables~9 and 10 on the other hand, had to be appraised by
examining the event's timing.

To accommodate cascade resonance, we next optimized the integration runs by
stepwise modifying the radial component of the separation velocity (and thereby
the perihelion and encounter times).  We continued to involve, in addition, the
normal, but not the transverse, component.  However, solutions were not optimized
to fit the comet's 1846 perihelion time.  Our aim was limited to investigating
the number of consecutive revolutions about the Sun over which the resonance
lock was in effect, and to find out whether the final orbital period, at the
time the lock broke down, fared well in comparison with the observed orbital
period of C/1846~O1 (Section~3.2).

We began by assuming that the fragmentation event took place 569~AU from the
Sun after aphelion, as in the early runs described in Section~4.  By extensively
varying the other parameters within tight limits, we tested the sensitivity
of the solutions to conditions of cascade resonance, a process that
consisted of hundreds of integration runs.  At the end of this stage of
experimentation we were able to bring the final orbital period down to 1819~yr,
which is inconsistent with a 3$\sigma$ limit ($\sim$1600~yr), obtained in
Section~3.2 from the comet's observations, by rather a narrow margin.
This solution, the first entry in Table~11, indicates that cascade resonance
unraveled after five returns to perihelion.  The integrated perturbations
$\Delta z_{\rm res}$ were found to be confined to a range from +0.000631 to
+0.000859~(AU)$^{-1}$, with{\vspace{-0.04cm}} a mean of
+0.000796$\:\pm\:$0.000110~(AU)$^{-1}$.

We next moved the parent comet's breakup to earlier times:\ first to a
heliocentric distance of 700~AU after aphelion, then to aphelion itself,
and then to three~pre\-aphelion locations.  These solutions, listed in
Table~11, offered steeper rates of inward drifting of aphelion than did
the 569~AU case.  The final orbital periods dropped below the 3$\sigma$
limit of $\sim$1600~yr to 1464~yr in the 700~AU post-aphelion case; to
1445~yr in the aphelion case; to 1414~yr in the 500 AU preaphelion case; and to
the shortest achieved period of 1353~yr in the 700 AU preaphelion case, in
which the integrated perturbations $\Delta z_{\rm res}$ ranged from~+0.000797
to{\vspace{-0.04cm}} +0.000969 (AU)$^{-1}\!$,~with~an~average of
\mbox{+0.000920$\:\pm\:$0.000072 (AU)$^{-1}$}.  The resonance unraveled
after 5--6 returns to perihelion in all scenarios except when the
fragmentation event occurred at 569 AU preaphelion and near the previous
perihelion.

Besides the orbits in Table 11, all derived for a total separation velocity of
$\sim$1~m~s$^{-1}$, further solutions were obtained for fragmentation at
700~AU preaphelion with a separation velocity of $\sim$2~m~s$^{-1}$, essentially
in the out-of-plane direction.  One of these runs resulted in a final orbital
period of 1231~yr, the shortest we found.  Presented as the second entry
in Table~12, its comparison with the first entry (copied from Table~11) confirms
a modest effect of the normal separation velocity on the rate of inward drifting.
As long as the resonance lock holds, the integrated perturbations $\Delta z_{\rm
res}$ in this high-velocity case range from +0.000841 to{\vspace{-0.03cm}}
+0.000995~(AU)$^{-1}$, averaging \mbox{+0.000940$\:\pm\:$0.000061 (AU)$^{-1}$}.
Even though this is only slightly higher than in the respective low-velocity case,
the corresponding difference in the final orbital period is seen to be more than
120~yr, or about 10\%.

\section{Remarks on Rapid Inward Drifting Scenarios}
To assess the significance of rapid inward drifting of aphelion for the evolution
of comets, it is desirable to address the proposed orbital-cascade resonance
process in terms of its orbit-changing power as well as from the standpoint
of its likelihood of occurrence among comets.

\subsection{Rates of Orbital-Period Change}
The most obvious limitation of the process is that only Jupiter qualifies
as a sufficiently effective perturber to allow this process to proceed.
Because of Jupiter's position in the solar system, the process applies only
to comets of orbital periods long enough that their aphelia are much more
than an order of magnitude greater than the Jovian distance from the Sun.
Crudely, a significant effect of this kind can be expected to manifest
itself only with comets whose initial orbital period substantially exceeds
$\sim$1000~yr.  Such comets can experience resonances with Jupiter of
1:$\Gamma$, where \mbox{$\Gamma \ge 100$}.  However, because the effect per
encounter is approximately constant in terms of the reciprocal semimajor
axis, $\Delta z_{\rm res}$, the respective rate of drop in the orbital
period, $\Delta P_{\rm res}$, per encounter depends strongly on the period
$P$ itself,
\begin{equation}
\Delta P_{\rm res} = - {\textstyle \frac{3}{2}} P^{\frac{5}{3}} \Delta
 z_{\rm res},  \;\;\;\;\;{\rm (for} \,\:\Delta P_{\rm res} \ll P).
\end{equation}
Accordingly, the process is most effective for comets in extremely elongated
orbits and, for a given comet, in the course of the first few revolutions
about the Sun after the initial encounter.  This circumstance immediately
leads to a notion that the process should be most effective for dynamically
new comets.  Indeed, if in the case we investigated in detail the comet
were initially arriving from the Oort cloud and had an orbital period of
$\sim$4~million yr, its period after the first Jovian encounter would be a
mere 30,000~yr, or less than 1\% of the original period.  This is truly
remarkable, given the encounters at rather common jovicentric distances of
$\sim$1~AU.  At the other extreme, comets with periods much shorter than
1000~yr could in the same situation drift inward at rates much lower than
150~yr per encounter.  Among known comets with periods shorter than 1000~yr
there is a group of about 20, whose Tisserand invariant with respect to
Jupiter is \mbox{$J < 2$}, most of them with periods between 60 and 200~yr,
that are sometimes referred to as Halley-type comets (Carusi et al.\ 1986).
Several of them were found to avoid close encounters with Jupiter, their
motions subject to libration patterns (Carusi et al.\ 1987a, 1987b).  The
librating comets were found to be in resonances of 1:5 to 1:7 with Jupiter,
but because of the range of the orbital periods, there is a potential for
resonances of up to 1:16, when investigated over a sufficiently long span
of time.

%
%
%
\begin{table}[t] 
\vspace{-3.8cm}
\hspace{4.21cm}
\centerline{
\scalebox{1}{
\includegraphics{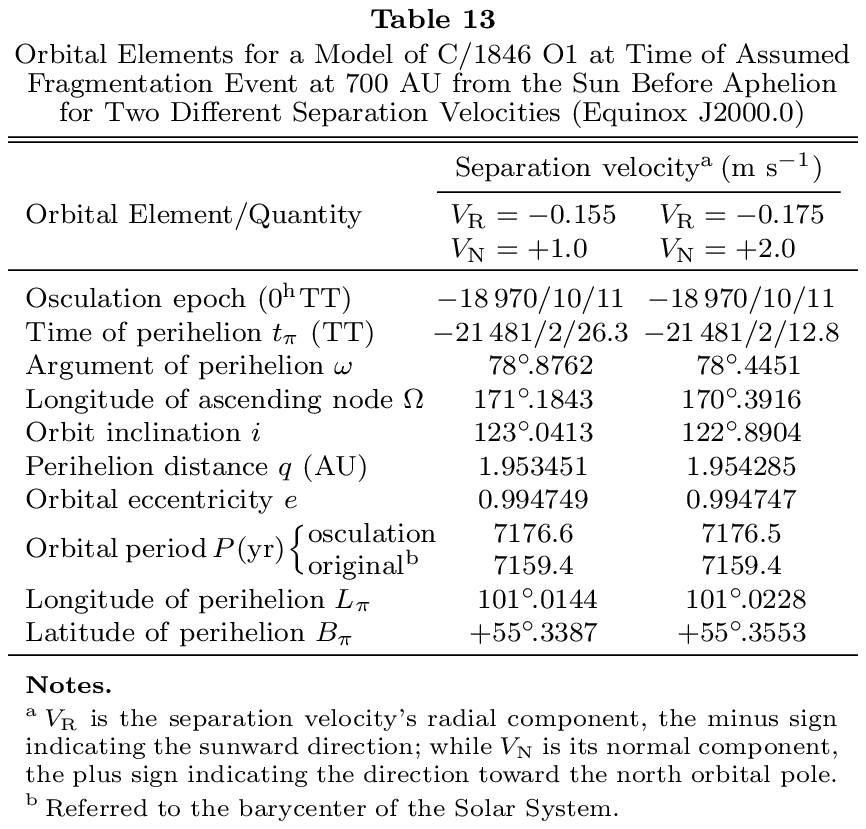}}} 
\vspace{-17.43cm}
\end{table}

\subsection{Verdict on the Pair of C/1846 O1 and C/1973 D1}

Returning now to our modeling of the inward drifting of C/1846~O1, we started
from the time of fragmentation event of the parent comet of the presumed pair
of C/1846~O1 and C/1973~D1, assumed to have occurred at some point of its orbit
between the perihelion time in $-$14\,326 and the previous perihelion passage,
in $-$21\,481.  The solutions that led to the shortest final orbital periods,
listed in Table~12, were based on the sets of osculation elements for C/1846~O1
at a selected location of the fragmentation event, 700~AU before aphelion, in
the year $-$18\,970.  For the two choices of the normal component of the
separation velocity, these elements are presented in Table~13.  Notable are
the large perihelion distances that dropped back to $\sim$1.4~AU by the time
of next perihelion, as shown in Table~14, in which we summarize the results
of orbit integration over 20--21 millennia.

\begin{table*}[t] 
\vspace{-3.8cm}
\hspace{-0.52cm}
\centerline{
\scalebox{1}{
\includegraphics{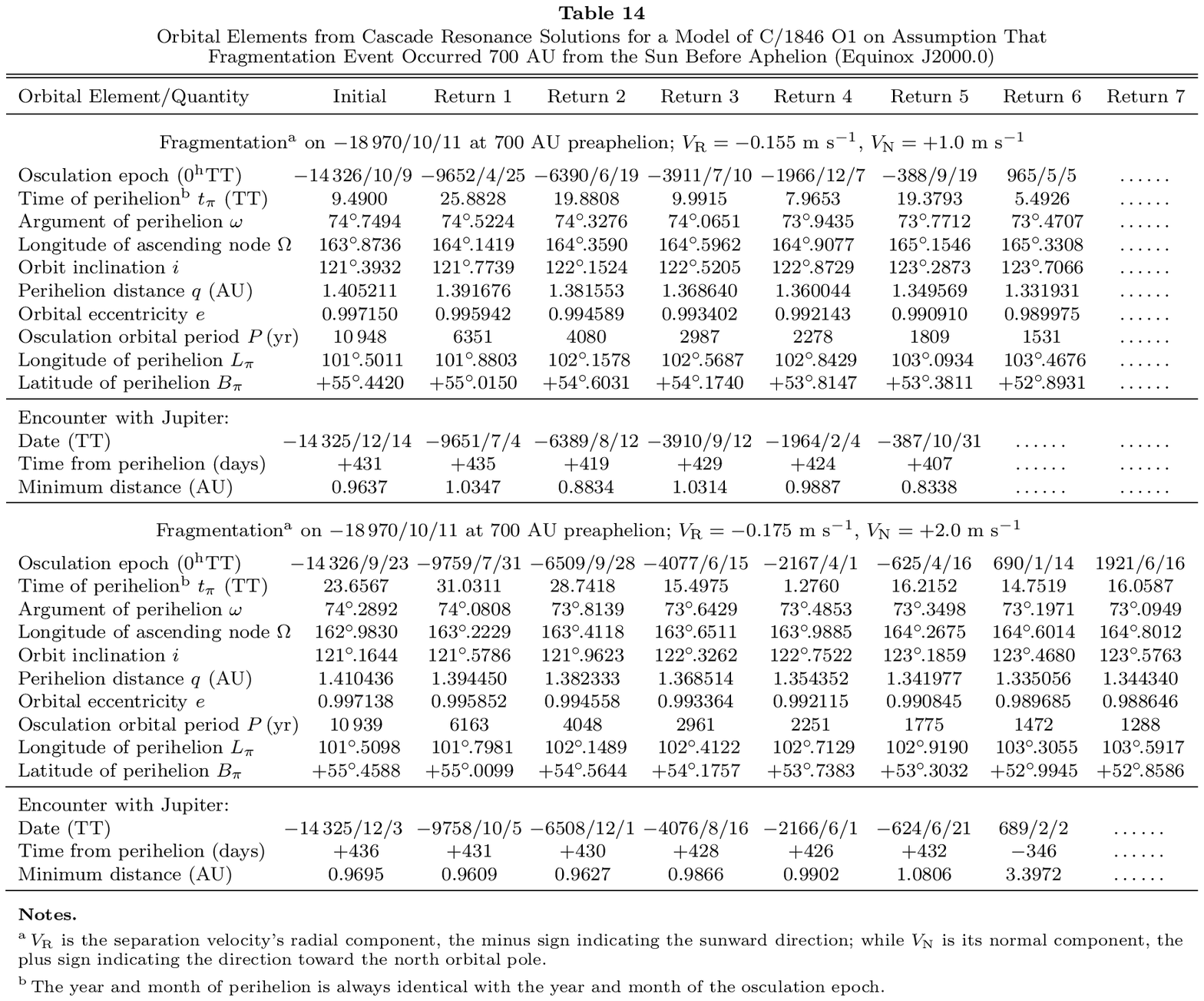}}} 
\vspace{-10.55cm}
\end{table*}

The orbital solutions in Tables~13 and 14 do unfortunately include effects
of chaotic motion owing to truncation in the computations.  In Section~3.5
we mentioned that after four encounters with Jupiter, a truncation error
accumulating over 20\,000~yr is no longer negligible even in high-precision
computations (17~digits), amounting to 1300~seconds.  In addition, truncation
errors increase exponentially with the number of encounters of approximately
equal perturbation effects.  Given that the error accumulated over
20\,000~yr with encounters absent is in a subsecond range, an expected
truncation error amounts to 0.12~days after five encounters, 1~day after
six encounters, and 8~days after seven encounters.

It therefore appears that the resonance lock broken after \mbox{5--7}~returns
to perihelion was not a product of orbital nature, but was brought on
computationally by truncation errors.  If so, one could expect that cascading
resonance might have continued.  Should this be the case, the results in
Tables~11--14 suggest that the orbital period of the modeled comet C/1846~O1
could eventually get below 1000~yr, but certainly not by 1846.

We thus arrive at a conclusion that it was the {\it much too slow rate of
inward aphelion drifting\/} caused by the inadequate Jovian perturbations,
{\it not\/} the {\it broken lock of cascade resonance\/}, that prevented the
final orbital period to drop below $\sim$1200~yr in our orbit integrations.
This explanation is supported by comparison with the computational schemes
listed in Tables~9--10.  While the genuine perturbations obtained by integrating
the motion of C/1846~O1 always remained, {\vspace{-0.03cm}}however slightly,
below \mbox{$\Delta z_{\rm res} = 0.001000$ (AU)$^{-1}$} per revolution, the
schematic sequences of perturbations that fitted the required rate of inward
drifting necessitated, in the least, the rates of \mbox{$\Delta z_{\rm res}
\approx 0.001200$ (AU)$^{-1}$} per revolution, but preferably \mbox{$\Delta
z_{\rm res} \geq 0.001700$ (AU)$^{-1}$} per revolution, in order for the orbital
period to drop, by 1846, below $\sim$1000~yr after the first four returns.  If,
as noted in Section~3.2, a 1$\sigma$ final orbital period of C/1846~O1 is less
than this limit, the examined scenario fails to explain this comet's evolution.

In addition, Table 14 shows that the long-term trends in the comet's other
elements are likewise unfavorable to the modeled scenario.  This is
particularly true~about the argument of perihelion, in which C/1846~O1 and
C/1973~D1 differ most significantly.  Whereas the observed value of this
angular element is by 2$^\circ\!$.5 greater for C/1846~O1, the model in Table~14
requires that it actually be smaller than that for C/1973~D1, deviating from the
expected value by about 4$^\circ$.  The results are also rather disappointing
in the other two angular elements, with only the perihelion distance being in
fair agreement with expectation.  Overall, these findings support a conclusion
that C/1846~O1 and C/1973~D1 either are {\it not genetically related\/} or
otherwise followed an evolutionary path {\it different\/} from the one we
proposed and examined; in Section~7 we briefly offer some speculations.
%

\subsection{Likelihood of{\vspace{-0.02cm}} Orbital-Cascade Resonance\\Among
Long-Period Comets} \vspace{-0.04cm}
To estimate the likelihood of a long-period comet getting locked into
orbital-cascade resonance, we employ a model of constant integrated
perturbation $\Delta z_{\rm res}$ introduced in Section~3.9.  This is
permissible because, on the one hand, the orbital computations leading
to Tables~11 and 12 show that the condition is approximately satisfied
in the course of the cascade-resonance process and, on the other hand,
comparison of Tables~9 and 10 with Tables~11 and 12, suggests common
similarities in terms of the deviations $\delta_{k,k+1}$ [defined by
Equation (15)] from exact resonance.  We employed these resonance
deviations from the 1:370 through 1:430 commensurabilities, that is
61 sets of model scenarios, for each of which we considered resonance
spanning 10 through 19 returns to perihelion.  Although these numbers
of returns are excessive, it turns out that there is no correlation
between the distribution of the deviations and the number of returns.
Besides, we were interested only in the deviations $\epsilon_4$ and
$\epsilon_5$, averaged, respectively, over the first four
{\nopagebreak}and five returns to perihelion, as defined by Equation
(11).{\vspace{-0.1cm}}
\begin{figure}[t] 
\vspace{0.15cm}
\hspace{-0.15cm}
\centerline{
\scalebox{0.8}{ 
\includegraphics{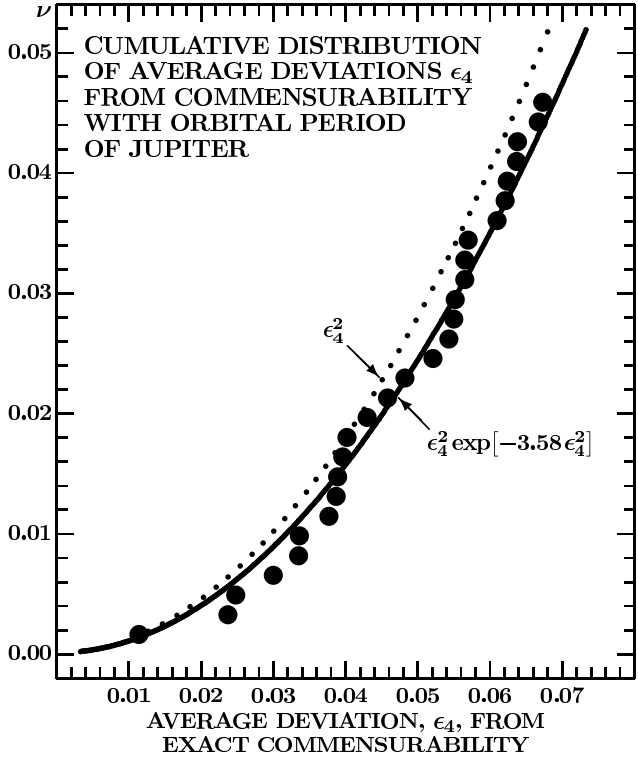}}} 
\vspace{-0.05cm}
\caption{Cumulative distribution $\nu_4$ of the averaged deviation $\epsilon_4$
from exact commensurability with the Jovian orbital period (1:370 through
1:430) after{\vspace{-0.05cm}} four consecutive returns.  The solid curve
is expressed by Equation (23), the dotted{\vspace{-0.095cm}} curve is an
$\epsilon_4^2$ approximation.  The distribution is normallized to
\mbox{$\nu_4(\frac{1}{2}) = 1$.}{\vspace{0.5cm}}}
\end{figure}

A cumulative distribution of the averaged deviations $\epsilon_4$ is
presented in Figure~6.  The plot shows clearly that in the range of small
deviations, which we are interested in, the cumulative distribution
increases with the square~of $\epsilon_4$.  From the definition it follows
{\vspace{-0.06cm}}that no deviation $\delta_{k,k+1}$ can exceed $\frac{1}{2}$,
so that $\nu(\epsilon_4)$, the cumulative distribution of $\epsilon_4$, may
be written in the form
\begin{equation}
\nu(\epsilon_4) = \nu_4 = A \, \epsilon_4^2 \, f(\epsilon_4),
\end{equation}
where $A$ is a constant, while $f(\epsilon_4)$ is a normalizing function that
has to satisfy a constraint
\begin{equation}
f({\textstyle \frac{1}{2}}) = \frac{4}{A}
\end{equation}
and two convergence conditions
\begin{equation}
\lim_{\epsilon_4 \rightarrow 0} f(\epsilon_4) = 1
\end{equation}
and, in order that \mbox{$d\nu_4/d\epsilon_4 > 0$} for any \mbox{$\epsilon_4
\leq \frac{1}{2}$},
\begin{equation}
\lim_{\epsilon_4 \rightarrow \frac{1}{2}} \frac{df(\epsilon_4)}{d\epsilon_4}
 > - \frac{16}{A}.
\end{equation}
It turns out that the slope $d(\ln \nu_4))/d(\ln \epsilon_4)$ must be
systematically decreasing with $\epsilon_4$ in order to satisfy these
conditions, given the data plotted in Figure 6.  We find that a Gaussian,
\begin{equation}
f(\epsilon_4) = \exp \left( -B \epsilon_4^2 \right),
\end{equation}
satisfies these conditions very well.  The condition that $\nu_4$ be an
increasing function requires a constraint \mbox{$B < 4$}, while the
normalization, \mbox{$\nu_4(\frac{1}{2}) = 1$}, imposes the following
relation between the constants $B$ and $A$:
\begin{equation}
B = 4 \ln \left( \frac{A}{4} \right).
\end{equation}
A satisfactory fit to the data points in Figure~6 is provided by a formula
\begin{eqnarray}
\nu_4 & = & 9.82\,\epsilon_4^2 \exp\!\left(-3.58 \, \epsilon_4^2
\right) \! . \nonumber \\[-0.1cm]
& & \mbox{\llap{$\pm$}} 0.83 \hspace{1.015cm} \pm \! 0.34
\end{eqnarray}

We also investigated the cumulative distribution $\nu(\epsilon_5)$ of an
averaged deviation from exact commensurability at the first five returns,
$\epsilon_5$, and found that the data obeyed the same type of law, namely,
\begin{eqnarray}
\nu(\epsilon_5) = \nu_5 & = & 5.02\,\epsilon_5^2 \exp\!\left(-0.91 \,
\epsilon_5^2 \right) \!. \nonumber \\[-0.1cm]
& & \mbox{\llap{$\pm$}} 0.51 \hspace{1.015cm} \pm \! 0.41
\end{eqnarray}
Somewhat surprisingly, at the same level of $\epsilon_m$, the probablility
of five consecutive near-resonance returns to peri\-helion is as high as
$\sim\!\!\frac{1}{2}$ the probability of four such consecutive returns.

We emphasize that the sequence of solutions introduced by Equation~(8) is
quite general, not limited to the circumstances of the modeled evolution of
C/1846~O1.  Numerically, we find from the eight runs listed in Tables~11 and
12 that an average $\epsilon_4$ is near 0.01, so once a comet gets perturbed
into a commensurable orbit upon its approach to Jupiter, the probability of
its motion getting locked into cascade resonance over four successive returns
to perihelion is about 10$^{-3}$ and over five successive returns about
\mbox{$0.5 \times 10^{-3}$}.

The probability of a temporary resonance lock is fairly low, but the likelihood
of detecting a comet subjected to this process is in fact higher, because --- as
its orbital period gets progressively shorter --- it enters, per unit time, the
perihelion region ever more often than a comet whose motion does not get locked
into cascade resonance.

\begin{figure}[t] 
\vspace{0.15cm}
\hspace{-0.15cm}
\centerline{
\scalebox{1.01}{
\includegraphics{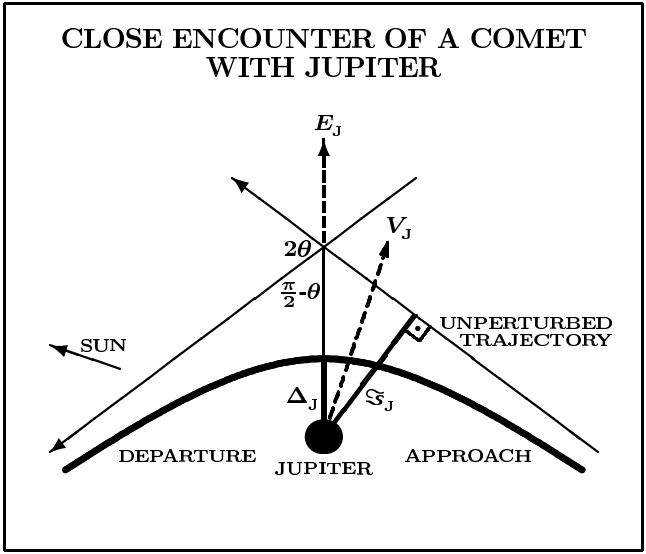}}} 
\vspace{0.02cm}
\caption{Schematic representation of the matched-conic approximation, applied
by Everhart (1969) in deriving an integrated perturbation $\Delta(1/a)$, per
revolution, as a result of a close encounter with Jupiter.  The comet moves
from the right to the left along the thick-drawn trajectory.  The angle between
the approach and departure trajectory branches is $2\theta$, the distance at
the time of closest approach to Jupiter is $\Delta_{\rm J}$, whereas $\Im_{\rm
J}$ is a minimum distance from Jupiter along the unperturbed trajectory (an
impact parameter), and {\boldmath $V_{\bf J}$} and {\boldmath $E_{\bf J}$} are,
respectively, Jupiter's orbital velocity vector and a unit vector along the
Jupiter--comet direction at perijove.{\vspace{0.3cm}}}
\end{figure}

\subsection{Integrated Perturbations Under Condition of\\Fairly Close
Encounter with Jupiter}
In Section 6.1 we expressed a conviction that, as a driver of rapid inward
drifting of aphelion, the process of high-order orbital resonance should --- in
the presence of regularly recurring fairly close encounters with Jupiter --- be
responsible for the existence of long-period comets with a short post-Oort-Cloud
history.  To place the rigorously determined Jovian perturbations integrated
over one revolution about the Sun in the context of the basic variables of the
perturbation theory, we write $\Delta(1/a)$, a perturbation of the reciprocal
semimajor axis integrated over a revolution and representing a total energy
change, as a function of the orbital elements (the eccentricity $e$, the
parameter $p$, and the orbital period $P$) and the radial and transverse
components of the perturbation acceleration, $\wp_{_{\rm R}}(t)$ and
$\wp_{_{\rm T}}(t)$, (e.g., Danby~1988):
\begin{equation}
\Delta(1/a) = -\frac{2\sqrt{p}}{k} \!\left[ \frac{e}{p} \int_{(P)} \!\!
 \wp_{_{\rm R}} \sin \upsilon \, dt + \!\! \int_{(P)} \!\!
\frac{\wp_{_{\rm T}}}{r} \, dt \right] \! ,{\vspace*{0.2cm}}
\end{equation}
where $k$ is the Gaussian gravitational constant, and the comet's orbital
position at time $t$ is defined by a heliocentric distance $r(t)$ and a true
anomaly $\upsilon(t)$.  Although the integration is carried out over an entire
revolution about the Sun, much of the total effect in the case of a fairly
close encounter comes from the perturbations along a rather short arc centered
on the point of closest approach.

Everhart (1969) investigated the problem of orbital perturbations as a result
of close encounters of comets, initially in parabolic orbits, with the planets
and showed that insight into the problem of an energy change is provided by a
matched-conic approximation,\footnote{In relation to the problem of earth-moon
trajectories in space missions, a matched-conic approximation was earlier
examined by Lagerstrom \& Kevorkian (1963).{\vspace{0.05cm}}} which uses
analytic expressions and offers results that agree with the results based on
the exact, numerical solution fairly satisfactorily.  This approximation is
shown schematically in Figure~7.  Everhart wrote the energy perturbation in
dimensionless units; using the absolute units, his expression for the $1/a$
perturbation integrated over an encounter with Jupiter, which is assumed to
move about the Sun in a circular orbit\footnote{The subsequently introduced
ellipticity of Jupiter's orbit was shown by Everhart (1972) to have no effect
on the rate of capture and the longitudinal distribution of perihelia.} of
radius $r_{\rm J}$, has the form~as follows:
\begin{equation}
\Delta(1/a) = \frac{4 V_{\rm rel} \sin \theta}{r_{\rm J} V_{\rm J}^2} \,
 (\mbox{\boldmath $V_{\bf J}$} \mbox{\boldmath $\cdot$} \mbox{\boldmath
 $E_{\bf J}$}),
\end{equation}
where $V_{\rm rel}$ is the magnitude of the asymptotic velocity of the comet
relative to Jupiter before the encounter (as well as after encounter, except
for the direction), {\boldmath $V_{\bf J}$} is the vector of Jupiter's orbital
velocity, whose magnitude is $V_{\rm J}$, {\boldmath $E_{\bf J}$} is a unit
vector from Jupiter to the comet at the time of closest approach, the dot
signifying a scalar product, and $2\theta$ is the angle by which the direction
of the departing branch of the comet's trajectory deviates from that of the
approaching trajectory (Figure 7).  Angle $\theta$ is related to the Jovian
mass, ${\cal M}_{\rm J}$, by
\begin{equation}
\tan \theta = \frac{{\cal M}_{\rm J}}{\mbox{\Msun}}\,\frac{r_{\rm J}}{\Im_{\rm J}}
 \,\frac{V_{\rm J}^2}{V_{\rm rel}^2},
\end{equation}
where {\Msun} is the Sun's mass and $\Im_{\rm J}$ is the distance of the
unperturbed approach trajectory from Jupiter, which Everhart calls an impact
parameter.  As the quantity of primary interest to us is the comet's distance
from Jupiter at the time of closest approach, $\Delta_{\rm J}$, we note that it
is related to $\Im_{\rm J}$ by
\begin{equation}
\Delta_{\rm J} = \Im_{\rm J} \, \frac{1 - \sin \theta}{\cos \theta}.
\end{equation}
After inserting for $\Im_{\rm J}$ from Equation (28) to (27) and then for $\theta$
from Equation~(27) to (26), we find that
\begin{equation}
\Delta(1/a) = 4 \, \frac{{\cal M}_{\rm J}}{\mbox{\Msun}} \, \frac{(\mbox{\boldmath
 $V_{\bf J} \! \cdot \! E_{\bf J}$})}{\Delta_{\rm J} V_{\rm rel} (1 \!+\! \Psi)},
\end{equation}
where
\begin{equation}
\Psi = \frac{{\cal M}_{\rm J}}{\mbox{\Msun}} \,\frac{r_{\rm J}}{\Delta_{\rm J}}
 \,\frac{V_{\rm J}^2}{V_{\rm rel}^2} = \frac{\sin \theta}{1 \!-\!\sin \theta}.
\end{equation}
In the limiting case, \mbox{$\lim_{\theta \rightarrow 0} \Psi \simeq \theta(1
\!+\! \theta)$}.

As an example of the degree of accuracy provided by Equation (29), we compare
the approximate values of $\Delta(1/a)$ with the numbers computed rigorously;
the latter of course include all perturbing bodies, not just Jupiter.  The
results, in Table~15, suggest that the approximation, expressed by Equation~(29),
yields $\Delta(1/a)$ values that are, on the average, \mbox{25$\:\pm\:$6} \%
higher than are the $\Delta z_{\rm res}$ values from the rigorous computations.

\begin{table*}[t] 
\vspace{-3.8cm}
\hspace{-0.5cm}
\centerline{
\scalebox{1}{
\includegraphics{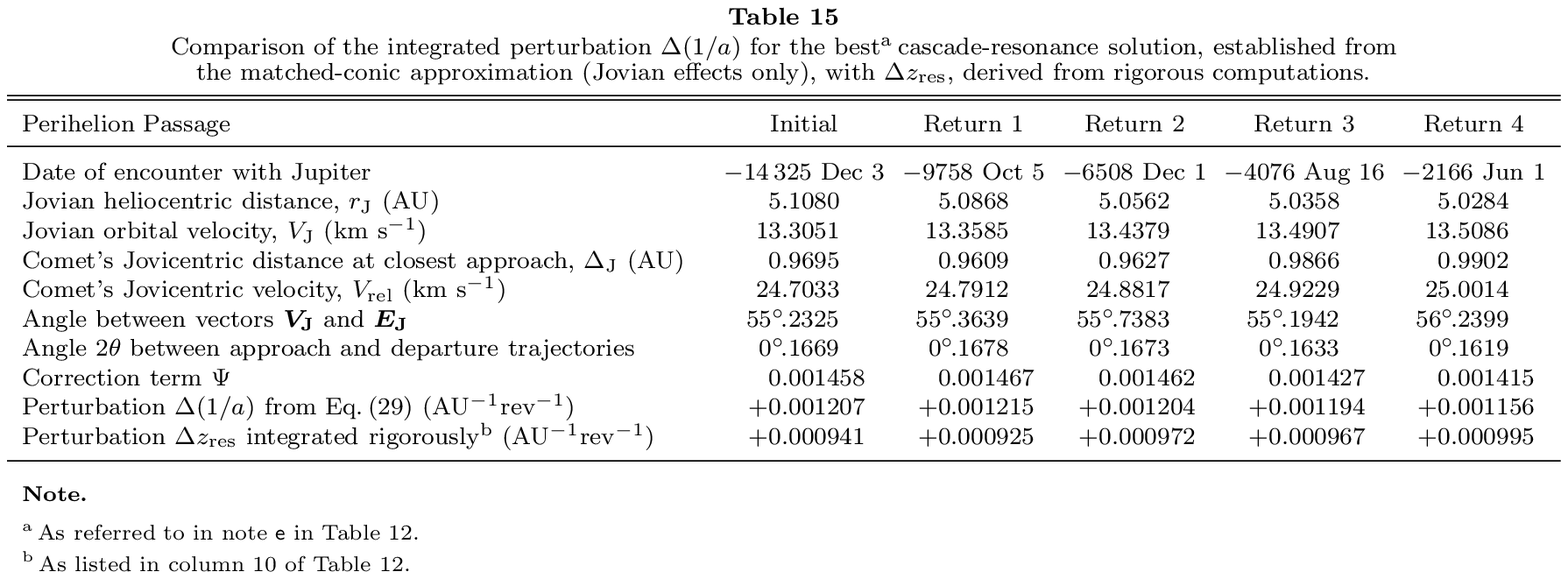}}} 
\vspace{-18.7cm}
\end{table*}

The value of $\Psi$ in Equation (29) can be perceived~as~a correction term,
{\vspace{-0.04cm}}for which we find \mbox{$\Psi \ll 1$} (and \mbox{$\theta \ll
30^\circ$}) when \mbox{$\Delta_{\rm J} \gg 0.005 \, (V_{\rm J}/V_{\rm rel})^2$},
that is, for nearly all encounters.  Equation~(29) then indicates that, excluding
cases of exceptionally close encounters, $\Delta(1/a)$ varies inversely as a
{\vspace{-0.04cm}}product of the distance at closest approach and the relative
velocity, \mbox{$\Delta(1/a) \,\, \mbox{\raisebox{0.3ex}{$\scriptstyle \propto$}}
\,\, \Delta_{\rm J}^{-1} V_{\rm rel}^{-1}$}, confirming the well-known facts
that the integrated perturbation is the greater the closer the approach is and
the slower the comet moves relative to Jupiter.

During Jovian encounters of comets, whose perihelion distances are much smaller
than the distance of Jupiter from the Sun, their motion is{\vspace{-0.04cm}}
nearly perpendicular to the planet's motion and their heliocentric velocities
$\sqrt{2}$ times higher, resulting {\vspace{-0.05cm}}in relative velocities that
can reasonably be approximated by \mbox{$V_{\rm rel} \simeq V_{\rm J}\sqrt{3}$}.
Taking now \mbox{$\Psi \rightarrow 0$}, inserting the numerical values of the
constants, and recognizing that \mbox{$|{\mbox{\boldmath $V_{\bf J} \!\cdot\!
E_{\bf J}$}}| \leq V_{\rm J}$}, the perturbation from Equation~(29) becomes in
absolute value
\begin{equation}
|\Delta(1/a)| \leq \frac{0.0022}{\Delta_{\rm J}} \;{\rm (AU)}^{-1}\, {\rm
rev}^{-1}.
\end{equation}
Given further that the averaging of the angle between~{\boldmath $V_{\bf J}$}
and {\boldmath $E_{\bf J}$} that is randomly distributed over $\pi/2$ results
in \mbox{$\langle | \mbox{\boldmath $V_{\bf J} \!\cdot\! E_{\bf J}$} | \rangle
= 2 V_{\rm J}/\pi$}, we find
\begin{equation}
\langle|\Delta(1/a)|\rangle \simeq \frac{0.0014}{\Delta_{\rm J}}\;{\rm (AU)}^{-1}\,
 {\rm rev}^{-1}.\\[0.05cm]
\end{equation}

Jovicentric velocities of comets whose perihelion distances are comparable to the
heliocentric distance of Jupiter, depend critically on the orbital inclination.
The relations (31) and (32) apply approximately when the inclination is near
90$^\circ$, but deviate from them increasingly as the inclination approaches
0$^\circ$ (prograde orbits) or 180$^\circ$ (retrograde orbits),{\vspace{-0.06cm}}
when the relative velocity converges to, respectively, \mbox{$V_{\rm rel} =
(\sqrt{2} \!-\! 1)V_{\rm J}$} and \mbox{$V_{\rm rel} = (\sqrt{2} \!+\! 1)V_{\rm
J}$}.  Relation (31) should for comets in essentialy coplanar orbits with
perihelion distances near Jupiter be replaced with
\begin{equation}
\Delta(1/a) \leq \frac{0.0092}{\Delta_{\rm J}}\;{\rm (AU)}^{-1}\,{\rm rev}^{-1}
\end{equation}
for prograde motions, and with
\begin{equation}
\Delta(1/a) \leq \frac{0.0016}{\Delta_{\rm J}}\;{\rm (AU)}^{-1}\,{\rm rev}^{-1}
\end{equation}
for retrograde motions.  For perihelia near Jupiter, the numerical coefficient
in relation (32) should likewise be higher for prograde orbits but lower for
retrograde orbits, compared to its value for small perihelion distances.

The retrograde-to-prograde ratio of 0.17 for cometary perihelia close
to Jupiter compares rather favorably with the respective results of two
independent investigations of 20\,000 hypothetical comets{\vspace{-0.035cm}}
in randomly distributed parabolic orbits, published by \v{S}teins \&
Kronkalne (1964) and by Fern\'andez (1981).  Likewise, the above relations
are generally consistent with the conclusion reached on the basis of more than
180\,000 hypothetical long-period comets by Everhart (1968), who found that
the $\Delta(1/a)$ perturbation as a function of perihelion distance depends
on the comet's inclination:\ in prograde orbits and orbits moderately
exceeding 90$^\circ$ a peak effect is attained at a perihelion distance
comparable with the orbital radius of the perturbing planet, whereas for
retrograde orbits an inconspicuously pronounced maximum takes place at
perihelion distances near zero.  Everhart's (1968) result averaged over all
inclinations shows a slight drop in the perturbations only at perihelion
distances smaller than $\sim$0.4~AU, but a systematic increase with increasing
perihelion distance by a factor of about 1.4 between 1 and 3~AU.  The peak is
reached at 5~AU, with a steep drop ensuing at \mbox{$r > r_{\rm J}$}.

In the context of the process of comet diffusion, the role of high-order
orbital-cascade resonance is reminiscent of very close encounters with
Jupiter --- the integrated perturbations in both cases trigger major changes
in the total orbital energy, except that the latter mode requires no
orbital-period commensurability.  Our computations for C/1846~O1 illustrate how
dramatic the similarity really is.  The peak total energy change, integrated
over all returns to perihelion (starting with those during which the orbital
motion was in the cascade-resonance lock), is represented by the second
solution in Tables~12 (referred to as the best-case scenario in footnote {\sf e})
and 14:\ the reduction of the orbital period from 7159.4~yr (Table 13) at the
fragmentation time to the final period of 1231.36~yr (Table~12) signifies an
integrated effect of~\mbox{$\Delta(1/a) = +0.008704 \!-\! 0.002692 = +0.006012$
(AU)$^{-1}$}, which --- with the values for the various
parameters~aver\-aged\footnote{Adopting{\vspace{-0.05cm}} \mbox{\mbox{$M_{\rm J}/
\mbox{\Msun} \!=\! 0.00095425$}, \mbox{$(\mbox{\boldmath $V_{\bf
J}\!\cdot\!E_{\bf J}$}) \!=\! 7.5909$ km\,s$^{-1}$},} \mbox{$V_{\rm
rel} = 24.8601$\,km\,s$^{-1}$}, and \mbox{$\Psi = 0.001446$}.{\vspace{0.03cm}}}
from the data listed in Table~15 --- is equivalent~to an effect of a single close
encounter with a minimum Jovicentric distance of \mbox{$\Delta_{\rm J} \simeq
0.194$ AU}.  From Sitarski's (1968) results it follows that only one long-period
comet observed between 1800 and the end of 1967 approached Jupiter to a similarly
small distance, namely, C/1932~P1 to 0.198~AU before perihelion.  The next three
ones were C/1840~E1, approaching the planet to 0.308~AU{\vspace{-0.04cm}}
(before perihelion); C/1823~Y1 to 0.448~AU (after perihelion);\footnote{A new,
elliptical {\vspace{-0.02cm}}orbit for C/1823 Y1 was computed by us in Paper~1,
implying that the distance of closest approach to Jupiter was 0.453~AU on 1824
Nov 14, in agreement with Sitarski's (1968) result, which is off by only
$-$0.005~AU and $-$1~day, respectively.} and C/1917~H1 to 0.489~AU (before
perihelion).

The rate of inward aphelion drifting in this best-case scenario is depicted in
Figure~8.  It shows this rate to decrease gradually from nearly 200~AU per
revolution after the initial perihelion passage in the year $-$14\,326 to
merely 10~AU per revolution between Returns 6 and 7 at the beginning of the 14th
century AD.
%
%

\begin{figure}[t] 
\vspace{0.12cm}
\hspace{-0.15cm}
\centerline{
\scalebox{0.8}{
\includegraphics{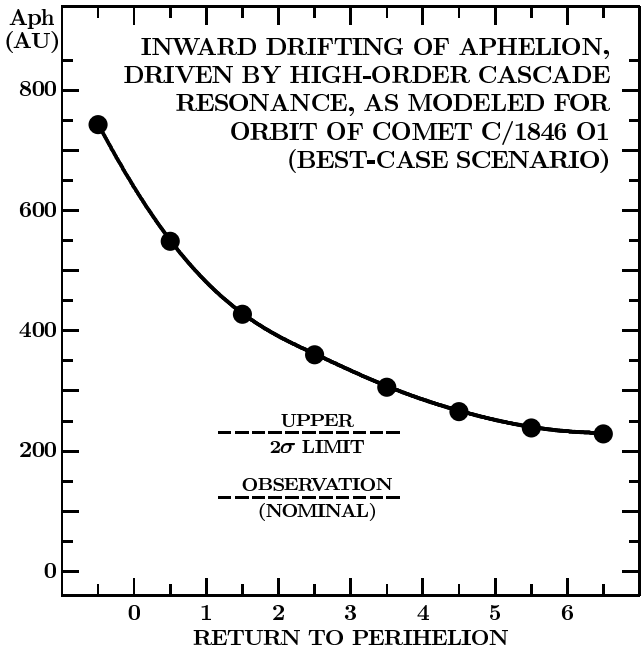}}} 
\vspace{0cm}
\caption{Computed rate of inward drifting of the aphelion distance, driven by
high-order orbital-cascade resonance, as function of the perihelion return.
This is the best-case scenario modeled for the motion of comet C/1846~O1.  The
first entry on the upper left is the aphelion distance after the fragmentation
event in the year $-$18\,970, but before the initial perihelion passage in the
year $-$14\,326 (Return 0).  The last entry on the lower right is the aphelion
distance between Returns 6 and 7 (Table~14), whose perihelion passages occurred,
respectively, in the years 690 and 1921.  The final aphelion distance is greater
than the nominal value derived from the observations (Table~4) by more than
100~AU, but it is very close to its upper 2$\sigma$ limit.{\vspace{0.5cm}}}
\end{figure}

Expanding now on the point discussed in Section 6.1, we confirm that, over an
interval of not more than a few dozen millenia, the process of high-order
orbital-cascade resonance could, under recurring Jovian approaches to $\sim$1~AU,
bring about changes in the orbital periods of the Oort-Cloud comets as prominent
as a single approach to about 0.2 AU of Jupiter could over a span of weeks or
months:\ for a comet with its perihelion much closer to the Sun than Jupiter,
the period could be reduced from $\sim$4~million~yr to $\sim$1700~yr, while in
the case of a near-perihelion encounter, to as little as $\sim$200~yr!
%
%
%

\section{Summary and Conclusions}
A statistical investigation of pairs and groups of long-period comets,
undertaken in terms of the Southworth-Hawkins $D$-criterion by Kres\'ak (1982),
resulted in a 20\% expectation for comets C/1846~O1 and C/1973~D1 making up a
random pair.  The expectation drops to~11\% when the highly inaccurate orbits
of the pre-19th-century comets are removed from the sample.  The degree of
ex\-pectation that would indicate the presence, in a random sample, of a nonrandom
event is essentially a matter of convention.  However, as a rule of thumb, a
lower-than-5\% random-event expectation is usually interpreted to mean that the
pair's members are genetically related, while a higher-than-25\% random-event
expectation suggests that their relationship is highly doubtful.  The degree
of expectation for the comet pair C/1846~O1 and C/1973~D1 is in a ``grey''
zone between the two limits, so their genetic relationship is uncertain.
Similarly, Lindblad (1985) found that the number of groups of long-period comets
varied with the $D$-criterion approximately as the upper 2$\sigma$ confidence
limit of the number of comet groups determined from searches in 20 random samples.

As a statistical tool, the $D$-criterion appears to be unfit for testing the
common origin of any specific comets.  Accordingly, we felt that it was desirable
to explore the potential relationship of the pair of C/1846~O1 and C/1973~D1 by
more rigorous techniques.

The first step was to reexamine the orbital motions of the two comets.  The
outcome, in Table~4, shows that our orbit for C/1973~D1 was remarkably similar to
that determined by Marsden et al.\ (1978), implying an original orbital period
of \mbox{16\,300$\:\pm\:$420}~yr.  On~the other hand, the currently cataloged
orbit for C/1846~O1 --- Vogel's (1868) set of parabolic elements with the
planetary perturbations unaccounted for --- was found to be unacceptable and in
need of major revision.  Our work revealed for the first time that the comet's
significant orbital property was a strong deviation from parabolic motion, with
a probable orbital period of $\sim$500~yr and not longer than $\sim$1600~yr
(3$\sigma$; Table~3).  This result unexpectedly complicated the problem of common
origin of the two comets by introducing a fundamental and previously unrecognized
stumbling block --- an enormous mismatch in their orbital periods.  However,
assuming a genetical relationship, the intrinsic brightness suggested that
C/1973~D1 should be a companion to C/1846~O1, while the orbital period of C/1973~D1
implied that the parent's splitting should have occurred more than $\sim$16\,000~yr
ago.

Integration of the orbital motion of C/1973 D1 back~in time showed that because of
an encounter with Jupiter, the planet's perturbations could either accelerate the
comet into an orbit of larger dimensions or decelerate it into an orbit of smaller
dimensions relative to those prior to the encounter, depending on the comet's
position with respect to the planet at the previous return in the year $-$14\,326.
A full range of orbital change achieved during one complete Jovian orbital period
(which equals merely 0.013 the rms uncertainty{\vspace{-0.03cm}} of the comet's
orbital period), amounts to 0.00196~(AU)$^{-1}$ in the barycentric recipriocal
semimajor axis, $1/a_{\rm b}$.  The motions of the two comets would have been
subjected to this differential effect if, in the year $-$14\,326, they passed
perihelion about 264~days, or almost 9~months, apart.  The perturbation curve
was far from symmetric, the maximum acceleration reaching $-$0.00113~(AU)$^{-1}$
and the {\vspace{-0.03cm}}maximum deceleration +0.00083~(AU)$^{-1}$.  If the
pre-encounter orbital periods were alike, C/1846~O1 must have been decelerated
during the Jovian encounter and the trailing C/1973~D1 accelerated.  Indeed,
the pre-encounter (original) orbital period of C/1973~D1 equaled 7160~yr
(Table~7),{\vspace{-0.02cm}} equivalent to \mbox{$(1/a_{\rm b})_{\rm orig} =
0.00156 + 0.00113 = +0.00269$}~(AU)$^{-1}$ and the approximate{\vspace{-0.03cm}}
post-encounter (future) orbital period of the modeled C/1846~O1{\vspace{-0.03cm}} 
was near 4800~yr,\footnote{This result is approximate because of the neglected
effect of a separation velocity at the time of fragmentation.} equivalent to
\mbox{$(1/a_{\rm b})_{\rm fut} = 0.00269 + 0.00083 = +0.00352$}~(AU)$^{-1}$,
in excellent agreement with the initial results of our rigorous computations
(Table~8).

These initial computations were based on a condition that the radial component
of the separation velocity between C/1846~O1 and C/1973~D1, which is the dominant
parameter determining the magnitude of the gap between the two objects at the
time of the $-$14\,326 perihelion, be exactly 1~m~s$^{-1}$.  In order for comet
C/1846~O1 to pass through perihelion first, it was required that the companion
(C/1973~D1) be released from the parent comet in the antisolar direction relative
to the primary fragment (C/1846~O1).\footnote{As it was the orbit of C/1973~D1
that, because of its considerably higher accuracy, was taken in our computations
as a reference orbit, this condition was implemented by releasing C/1846~O1 toward
the Sun relative to C/1973~D1.{\vspace{0.04cm}}}  The 1~m~s$^{-1}$ constraint was
satisfied by assigning the fragmentation event a time some 1470~yr before the
$-$14\,326 perihelion, when the parent was at a heliocentric distance of 569~AU
along the incoming leg of the orbit (i.e., post-aphelion).  In this scenario, the
modeled C/1846~O1 approached Jupiter to 0.99~AU about 430~days after perihelion,
in $-$14\,325 December (Table~8), while C/1973~D1 approached the planet to 0.87~AU
about 600~days after perihelion, in $-$14\,324 early June.

The early forward-integration runs for the modeled motion of C/1846 O1 beyond
the year $-$14\,326 showed that the initial post-fragmentation orbital period of
this object was equal to slightly less than 4800~yr (Table~8).  Even though this
is dramatically (by a factor of more than 3) shorter than the orbital period
of C/1973~D1 (4800~yr vs 16\,300~yr), it was deemed requisite that the period
of C/1846~O1 be further reduced by a factor of \mbox{3--4} to bring it to at
least a modest agreement with the orbital period of C/1846~O1 derived from the
observations (Sec.~3.2); a near-perfect correspondence would require a factor
of close to 10.

In an effort to slash the comet's consumed orbital time as much as possible, we
examined the dependence of the initial post-fragmentation orbital period
(between the year $-$14\,326 and Return~1) on the normal component of the
separation velocity and on the fragmentation time.  As shown in Tables~8 and 12,
a normal component of the separation velocity\footnote{The positive sign of the
velocity in Tables~8 and 12 means that the companion C/1973~D1 was, relative to
the primary C/1846~O1, released to the south of the parent's orbital plane.}
of \mbox{1--2}~m~s$^{-1}$ led to a slightly shorter initial post-fragmentation
orbital period (by \mbox{1--2}\%).  The shortest initial periods were obtained
for a fragmentation event taking place $\sim$700~AU from the Sun before aphelion,
in the year $-$18\,970 (or 4644~yr before the $-$14\,326 perihelion), the
minimum initial post-fragmentation period amounting then to $<$4600~yr (Table~12),
when the unlikely solution for fragmentation near the previous perihelion in
the 22st millennium BCE (Table~11) is ignored.

The described rationale not only resulted in a substantial reduction in the initial
post-fragmentation orbital period of the modeled motion of C/1846~O1, but it also
accentuated the merit of the applied numerical experiment, which aimed --- by
exploiting the perturbation effects during the recurring moderate encounters with
Jupiter --- at repetitively slashing the orbital period over several consecutive
revolutions about the Sun.  The urgent need for an additional reduction of the
orbital period of C/1846~O1 was illustrated by the factor of more than 9 between
the shortest initial post-fragmentation period of $\sim$4600~yr and the ultimate
target of $\sim$500~yr (Table~4).  The clue to this experiment's success was to
temporarily lock the comet's motion into high-order orbital-cascade resonance, a
process in which the aphelion was rapidly drifting inward over several consecutive
revolutions about the Sun, while the orbital period was successively equal to
gradually decreasing integral multiples of the Jovian orbital period.

The solution that we refer to as the best-case scenario offered for C/1846~O1 an
initial post-fragmentation orbital period of 4567~yr (Table~12) and resulted ---
after seven returns to perihelion (even though over only first five returns was
the orbital period in a resonance lock with Jupiter's period in a ratio of,
respectively, 1:385, 1:274, 1:205, 1:161, and 1:130) --- in the comet's final
orbital period of 1231~yr, the shortest final period that we were able to come up
with.  This period is shorter than the estimated 3$\sigma$ upper limit (Table~3)
to, but still nearly 2.5 times longer than, the most probable period for
C/1846~O1 listed in Table~4.  Because of the additional problems (such as
a discrepancy of about 4$^\circ$ in the comet's argument of perihelion),
we concluded that C/1846~O1 and C/1973~D1 were either genetically unrelated or
related in a different manner.  In any case, the process of high-order
orbital-cascade resonance did not explain the required rate of inward aphelion
drifting to complete satisfaction.  Our computations showed that in the modeled
motion of C/1846~O1 the peak perturbation rate, integrated over one revolution,
of the reciprocal semimajor axis was close to 0.0010~(AU)$^{-1}$, whereas the
rate, needed to explain the final orbital period in a general range of, say,
\mbox{500--800}~yr after not more than five returns to perihelion, was near
0.0017~(AU)$^{-1}$; such a rate would require [Equation~(27)] a minimum
distance of unachievable $\sim$0.5--0.6~AU rather than $\sim$0.9--1.0~AU
(Tables~11--12) some 440 days after perihelion, at the critical time during
the recurring encounters with Jupiter.

Comets C/1846 O1 and C/1973 D1 still could be genetically related, but their
evolutionary paths must have differed from the examined scenario in one way or
another (e.g., the parent's fragmentation event might have taken place one or
more revolutions earlier).  In a speculative example one may suggest a similar
process, but accompanied in addition by a series of fragmentation events involving
C/1846~O1 between the perihelion passages in the years $-$14\,326 and 1846.
While the resulting companions would probably fail to survive, the orbital
motion of the primary would~in each such event be affected by a momentum change,
arriving at Jupiter at a slightly different time.  As Jovian encounter-driven
perturbations are highly time-sensitive and capable of magnifying the slight
positional changes into more significant orbital-energy changes, the rate of
inward aphelion drifting might increase appreciably and the orbital period
shorten accordingly.  A similar outcome may also be triggered by a
nongravitational acceleration that could affect the comet's orbital motion.
Other than that, we do not see a competing tractable interpretation~of~the~two
long-period comets as a genetically related pair.

Our comprehensive investigation of high-order orbital-cascade resonance has
ramifications for the evolution of some comets.  We remark in passing that the
general problem of orbital perturbations in the presence of multiple encounters
with a planet was touched upon --- but dismissed as hardly tractable --- by
Everhart (1969) in connection with the now-solved problem of capturing comets
into short-period orbits.  Although the likelihood of orbital-cascade resonance
among comets is relatively low, the process represents an attractive mechanism for
a rapid rate of inward drifting of aphelia of long-period comets in general and
of aphelia of the dynamically new, Oort-Cloud comets in particular.  In the
context of the process of comet diffusion, the ultimate changes in the total
orbital energy of comets subjected to Jovian-driven high-order cascade resonance
exemplify effects that in terms of orbit-transformation severity compare
favorably to effects triggered in the course of very close encounters with
Jupiter, except that the latter do not require the orbital-period commensurability
and happen instantly.\\[-0.2cm]

This research was carried out in part at the Jet Propulsion Laboratory, California
Institute of Technology, under contract with the National Aeronautics and Space
Administration.

\begin{center}
{\footnotesize REFERENCES}
\end{center}
\vspace*{-0.4cm}
\begin{description}
{\footnotesize
\item[\hspace{-0.3cm}]
Argelander, F. W. A. 1865, AN, 63, 285
\\[-0.57cm]
%
%
\item[\hspace{-0.3cm}]
Bishop, G. 1852, Astronomical Observations Taken at the Obser-{\linebreak}
{\hspace*{-0.6cm}}vatory South Villa, Inner Circle, Regent's Park, London,
During{\linebreak}
{\hspace*{-0.6cm}}the Years 1839--1851. (London:\ Taylor, Walton, \&
Maberly), 215
\\[-0.57cm]
\item[\hspace{-0.3cm}]
Bortle, J. E. 1982, Intern. Comet Quart., 4, 79
\\[-0.57cm]
\item[\hspace{-0.3cm}]
Callandreau, O. 1892, Ann. Obs. Paris, 20, B.57
\\[-0.57cm]
\item[\hspace{-0.3cm}]
Carusi, A., Kres\'ak, L., Perozzi, E., \& Valsecchi, G. B. 1985, Long-{\linebreak}
{\hspace*{-0.6cm}}Term Evolution of Short-Period Comets. (Bristol, UK: A. Hilger
{\hspace*{-0.6cm}}Ltd.)
\\[-0.57cm]
\item[\hspace{-0.3cm}]
Carusi, A., Kres\'ak, L., Perozzi, E., \& Valsecchi, G. B. 1986, in{\linebreak}
{\hspace*{-0.6cm}}Exploration of Halley's Comet, ESA SP-250, ed. B. Battrick,{\linebreak}
{\hspace*{-0.6cm}}E. J. Rolfe, \& R. Reinhard (Nordwijk, Netherlands: ESTEC),{\linebreak}
{\hspace*{-0.6cm}}vol. 2, 413
\\[-0.57cm]
\item[\hspace{-0.3cm}]
Carusi, A., Kres\'ak, L., Perozzi, E., \& Valsecchi, G. B. 1987a,{\linebreak}
{\hspace*{-0.6cm}}A\&A, 187, 899
\\[-0.57cm]
\item[\hspace{-0.3cm}]
Carusi, A., Valsecchi, G. B., Kres\'ak, L., \& Perozzi, E. 1987b,{\linebreak}
{\hspace*{-0.6cm}}Publ. Astron.  Inst. Czech. Acad. Sci., 67, 29
\\[-0.57cm]
\item[\hspace{-0.3cm}]
Combi, M. R., DiSanti, M. A., \& Fink, U. 1997, Icarus, 130, 336
\\[-0.57cm]
\item[\hspace{-0.3cm}]
Danby, J. M. A. 1988, Fundamentals of Celestial Mechanics, 2nd{\linebreak}
{\hspace*{-0.6cm}}ed. (Richmond, VA: Willmann-Bell Publ.), 466pp
\\[-0.57cm]
\item[\hspace{-0.3cm}]
de Vico, F. 1846, Comptes Rendus Acad. Sci. Paris, 23, 477
\\[-0.57cm]
\item[\hspace{-0.3cm}]
Dones, L., Weissman, P. R., Levison, H. F., \& Duncan, M. J. 2004,{\linebreak}
{\hspace*{-0.6cm}}in Comets II, ed.\ M. C. Festou, H. U. Keller, \& H. A. Weaver
{\hspace*{-0.6cm}}(Tucson, AZ: University of Arizona), 153
\\[-0.57cm]
\item[\hspace{-0.3cm}]
Duncan, M., Levison, H., \& Dones, L. 2004, in Comets II, ed.{\linebreak}
{\hspace*{-0.6cm}}M. C. Festou, H. U. Keller, \& H. A. Weaver
(Tucson,~AZ:~Univer-{\linebreak}
{\hspace*{-0.6cm}}sity of Arizona), 193
\\[-0.57cm]
\item[\hspace{-0.3cm}]
Everhart, E. 1968, AJ, 73, 1039
\\[-0.57cm]
\item[\hspace{-0.3cm}]
Everhart, E. 1969, AJ, 74, 735
\\[-0.57cm]
\item[\hspace{-0.3cm}]
Everhart, E. 1972, in The Motion, Evolution of Orbits, and Origin{\linebreak}
{\hspace*{-0.6cm}}of Comets, Proc.\ IAU Symp. 45, ed.\ G. A. Chebotarev,~E.~I.
{\hspace*{-0.6cm}}Kazimirchak-Polonskaya, \& B. G. Marsden (Dordrecht, Nether-
{\hspace*{-0.6cm}}lands: Reidel Publ.), 360
\\[-0.57cm]
\item[\hspace{-0.3cm}]
Fern\'andez, J. A. 1981, A\&A, 96, 26
\\[-0.57cm]
\item[\hspace{-0.3cm}]
Fern\'andez, J. A., \& Gallardo, T. 1994, A\&A, 281, 911
\\[-0.57cm]
\item[\hspace{-0.3cm}]
Graham, A. 1846, MNRAS, 7, 160 \\[-0.57cm]
\item[\hspace{-0.3cm}]
Hind, J. R. 1846, AN, 24, 325
\\[-0.37cm]
\pagebreak
%
%
\item[\hspace{-0.3cm}]
Kazimirchak-Polonskaya, E. I. 1967, Soviet Astron., 11, 349
\\[0.27cm]
\item[\hspace{-0.3cm}]
Kazimirchak-Polonskaya, E. I. 1972, in The Motion, Evolution~of{\linebreak}
{\hspace*{-0.6cm}}Orbits, and Origin of Comets, IAU Symp.\,45, ed. G.\,A.\,Chebota-{\linebreak}
{\hspace*{-0.6cm}}rev, E. I. Kazimirchak-Polonskaya, \& B. G. Marsden (Dordrecht,
{\hspace*{-0.6cm}}Netherlands: Reidel), 373
\\[-0.57cm]
\item[\hspace{-0.3cm}]
Kohoutek, L. 1973a, IAU Circ., 2504
\\[-0.57cm]
\item[\hspace{-0.3cm}]
Kohoutek, L. 1973b, IAU Circ., 2593
\\[-0.57cm]
\item[\hspace{-0.3cm}]
Kojima, N. 1973, IAU Circ., 2511
\\[-0.57cm]
%
%
\item[\hspace{-0.3cm}]
Kres\'ak, L.$\!\!$' 1982, Bull. Astron. Inst. Czech., 33, 150
\\[-0.57cm]
\item[\hspace{-0.3cm}]
Lagerstrom, P. A., \& Kevorkian, J. 1963, AJ, 68, 84
\\[-0.57cm]
\item[\hspace{-0.3cm}]
Le Verrier, U.-J. 1857, Ann. Obs. Imp. Paris, 3, 203
\\[-0.57cm]
%
%
\item[\hspace{-0.3cm}]
L\'evy, P. 1937, Th\'eorie de l'Addition des Variables Al\'eatoires.{\linebreak}
{\hspace*{-0.6cm}}(Paris: Gauthier-Villiers).
\\[-0.57cm]
\item[\hspace{-0.3cm}]
Lexell, A. I. 1778, Acta Acad. Petropol., 1, 317
\\[-0.57cm]
\item[\hspace{-0.3cm}]
Lindblad, B. A. 1985, in Dynamics of Comets: Their Origin and{\linebreak}
{\hspace*{-0.6cm}}Evolution, Proc.  IAU Coll.\ 83, ed.\ A. Carusi \& G. B.
Valsecchi{\linebreak}
{\hspace*{-0.6cm}}(Dordrecht, Netherlands: Reidel Publ.), 353
\\[-0.57cm]
%
%
\item[\hspace{-0.3cm}]
Marsden, B. G. 1967, AJ, 72, 1170
\\[-0.57cm]
\item[\hspace{-0.3cm}]
Marsden, B. G. 1973, IAU Circ., 2504
\\[-0.57cm]
\item[\hspace{-0.3cm}]
Marsden, B. G. 1974, IAU Circ., 2646
\\[-0.57cm]
\item[\hspace{-0.3cm}]
Marsden, B. G. 1979, Catalogue of Cometary Orbits, 3rd ed.{\linebreak}
{\hspace*{-0.6cm}}(Cambridge, MA:\ Smithsonian Astrophysical Observatory)
\\[-0.57cm]
\item[\hspace{-0.3cm}]
Marsden, B. G. 1989, AJ, 98, 2306
\\[-0.57cm]
\item[\hspace{-0.3cm}]
Marsden, B. G., \& Williams, G. V. 2008, Catalogue of Cometary{\linebreak}
{\hspace*{-0.6cm}}Orbits 2008, 17th ed. (Cambridge, MA:\
Smithsonian Astrophysi-{\linebreak}
{\hspace*{-0.6cm}}cal Observatory)
\\[-0.57cm]
\item[\hspace{-0.3cm}]
Marsden, B.\,G., Sekanina, Z., \& Yeomans, D.\,K.\ 1973, AJ,\,78,\,211
\\[-0.57cm]
\item[\hspace{-0.3cm}]
Marsden, B. G., Sekanina, Z., \& Everhart, E. 1978, AJ, 83, 64
\\[-0.57cm]
\item[\hspace{-0.3cm}]
Morbidelli, A., \& Brown, M. E. 2004, in Comets II, ed.\ M. C.{\linebreak}
{\hspace*{-0.6cm}}Festou, H. U.  Keller, \& H. A. Weaver (Tucson, AZ: University
{\hspace*{-0.6cm}}of Arizona), 175
\\[-0.57cm]
%
%
%
\item[\hspace{-0.3cm}]
Morris, C. S. 1973, PASP, 85, 470
\\[-0.57cm]
%
%
\item[\hspace{-0.3cm}]
\"{O}pik, E. J. 1966, Irish Astron. J., 7, 141
\\[-0.57cm]
\item[\hspace{-0.3cm}]
\"{O}pik, E. J. 1971, Irish Astron. J., 10, 35
\\[-0.57cm]
\item[\hspace{-0.3cm}]
Rickman, H. 2004, in Comets II, ed. M. C. Festou, H. U. Keller,~\&{\linebreak}
{\hspace*{-0.6cm}}H. A. Weaver (Tucson, AZ: University of Arizona), 205
\\[-0.57cm]
\item[\hspace{-0.3cm}]
Sekanina, Z. 1982, in Comets, ed. L. L. Wilkening (Tucson, AZ:
{\hspace*{-0.6cm}}University of Arizona), 251
\\[-0.57cm]
\item[\hspace{-0.3cm}]
Sekanina, Z., \& Chodas, P. W. 2004, ApJ, 607, 620
\\[-0.57cm]
\item[\hspace{-0.3cm}]
Sekanina, Z., \& Kracht, R. 2016, ApJ, 823, 2 (Paper 1)
\\[-0.57cm]
\item[\hspace{-0.3cm}]
Sekanina, Z., \& Yeomans, D. K. 1984, AJ, 89, 154
\\[-0.57cm]
\item[\hspace{-0.3cm}]
Shlesinger, M. F., West, B. J., \& Klafter, J.\,1987, Phys. Rev. Lett.,{\linebreak}
{\hspace*{-0.6cm}}58, 1100
\\[-0.57cm]
\item[\hspace{-0.3cm}]
Sitarski, G. 1968, Acta Astron., 18, 171
\\[-0.57cm]
%
%
\item[\hspace{-0.3cm}]
Southworth, R. B., \& Hawkins, G. S. 1963, Smithson.\ Contr.{\linebreak}
{\hspace*{-0.6cm}}Astrophys., 7, 261
\\[-0.57cm]
\item[\hspace{-0.3cm}]
\v{S}teins, K., \& Kronkalne, S. 1964, Acta Astron., 14, 311
\\[-0.57cm]
\item[\hspace{-0.3cm}]
van Woerkom, A. J. J. 1948, BAN, 10, 445
\\[-0.57cm]
\item[\hspace{-0.3cm}]
Vogel, H. 1868, AN, 71, 97
\\[-0.57cm]
\item[\hspace{-0.3cm}]
Vsekhsvyatsky, S. K. 1958, Fizicheskie kharakteristiki komet{\linebreak}
{\hspace*{-0.6cm}}(Moscow:\ Gosud.\,izd-vo fiz.-mat.\,lit.); translated:\ 1964,
Physical{\linebreak}
{\hspace*{-0.6cm}}Characteristics of Comets, NASA TT-F-80 
(Jerusalem:~Israel
{\hspace*{-0.6cm}}Program for Scientific Translations)
\\[-0.57cm]
\item[\hspace{-0.3cm}]
Whipple, F. L. 1977, Icarus, 30, 736
\\[-0.67cm]
\item[\hspace{-0.3cm}]
Zhou, J.-L., Sun, Y.-S., \& Zhou, L.-Y. 2002,{\vspace{-0.1cm}} Cel.\ Mech.\ \&
Dyn.{\linebreak}
{\hspace*{-0.6cm}}Astron., 84, 409}
\vspace*{0.15cm}
\end{description}
\end{document}